\documentclass[11pt]{article}

\usepackage[preprint]{acl}

\usepackage{times}
\usepackage{latexsym}

\usepackage[T1]{fontenc}

\usepackage[utf8]{inputenc}

\usepackage{microtype}

\usepackage{inconsolata}

\usepackage{graphicx}

\usepackage{amsmath,amsfonts,bm}

\def\eqref#1{equation~\ref{#1}}

\def\1{\bm{1}}

\DeclareMathAlphabet{\mathsfit}{\encodingdefault}{\sfdefault}{m}{sl}
\SetMathAlphabet{\mathsfit}{bold}{\encodingdefault}{\sfdefault}{bx}{n}

\usepackage{tcolorbox}
\usepackage{hyperref}
\usepackage{url}
\usepackage{graphicx}
\usepackage{booktabs}
\usepackage{tabularx}
\usepackage{siunitx}
\usepackage{makecell}   
\usepackage{multirow}   
\usepackage{amssymb}
\usepackage{amsmath,amssymb} 
\usepackage{algorithm}
\usepackage{algpseudocode}
\algtext*{EndWhile}
\algtext*{EndIf}
\algtext*{EndFor}
\usepackage{tikz}
\newcommand{\circnum}[1]{%
  \tikz[baseline=(char.base)]{
    \node[shape=circle, draw=black, fill=black,
          inner sep=1.2pt, text=white, font=\footnotesize\bfseries] (char) {#1};
  }%
}
\usepackage{enumitem}

\usepackage{subcaption}
\usepackage[table]{xcolor}

\usetikzlibrary{arrows.meta, positioning, shapes.multipart, calc}

\newcolumntype{N}{S[table-format=2.1]}                
\newcolumntype{A}{S[table-format=2.1,parse-numbers=false]} 
\newcommand{\posdelta}[1]{\scriptsize\textcolor{gray}{#1}}
\newcommand{\negdelta}[1]{\scriptsize\textcolor{gray}{#1}}

\newcommand{\methodname}{\textsc{SParC-RAG}} 

\usepackage{xcolor}

\def\BibTeX{{\rm B\kern-.05em{\sc i\kern-.025em b}\kern-.08em
    T\kern-.1667em\lower.7ex\hbox{E}\kern-.125emX}}

\definecolor{darkgreen}{rgb}{0.0, 0.5, 0.0}
\definecolor{darkblue}{RGB}{31, 119, 180}

\definecolor{PaleGreen}{HTML}{E8FAE8}
\definecolor{PaleRed}{HTML}{FFEBEB}     
\definecolor{PaleGrey}{HTML}{F5F7FA}
\definecolor{PosCellGreen}{HTML}{C6F7C6}

\definecolor{DIblue1}{HTML}{5393D5} 
\definecolor{DIblue2}{HTML}{7EA9E1}
\definecolor{DIblue3}{HTML}{A6C8ED}
\definecolor{DIblue4}{HTML}{D0E3F7}
\definecolor{DIblue5}{HTML}{EAF2FB} 
\definecolor{Green0}{HTML}{F7FCF9} 
\definecolor{Green1}{HTML}{E9F9ED} 
\definecolor{Green1.5}{HTML}{CDEFD7} 
\definecolor{Green2}{HTML}{B7E4C7} 
\definecolor{Green2.5}{HTML}{86C89C}
\definecolor{Green3}{HTML}{51B37F} 

\definecolor{customcyan}{RGB}{0, 158, 115} 
\definecolor{tealblue}{RGB}{0, 114, 178}
\definecolor{darkorange}{RGB}{213, 94, 0}

\usepackage{setspace}
\tcbuselibrary{skins,breakable}

\newtcolorbox{promptbox}[2][]{%
  enhanced,
  width=0.95\textwidth,
  before skip=10pt,
  after skip=10pt,
  colback=gray!3,
  colframe=black,
  boxrule=0.6pt,
  arc=2pt,
  left=6pt,
  right=6pt,
  top=12pt,
  bottom=6pt,
  colbacktitle=black,
  coltitle=white,
  fonttitle=\normalsize\bfseries,
  title={#2},
  attach boxed title to top left={yshift=-\tcboxedtitleheight/2, xshift=2mm},
  boxed title style={boxrule=0.6pt, colback=black, colframe=black, rounded corners},
  before upper={%
    \begingroup
    \setstretch{1.1}
    \ttfamily
    \fontsize{9}{11}\selectfont 
    \raggedright
    \parindent=0pt
    \parskip=0pt
    \obeylines   
  },
  after upper={%
    \endgroup
  },
  #1
}

\definecolor{BestBg}{HTML}{CCE6F8}    
\definecolor{SecondBg}{HTML}{EDF9FC}  

\newcommand{\best}[1]{\cellcolor{BestBg}{\ensuremath{#1}}}
\newcommand{\secondbest}[1]{\cellcolor{SecondBg}{\ensuremath{#1}}}

\DeclareRobustCommand{\bestlegend}{%
  \begingroup
  \setlength{\fboxsep}{1pt}%
  \colorbox{BestBg}{%
    \rule{0pt}{1.2ex}
    \smash{\footnotesize Best}%
  }%
  \endgroup
}

\DeclareRobustCommand{\secondlegend}{%
  \begingroup
  \setlength{\fboxsep}{1pt}%
  \colorbox{SecondBg}{%
    \rule{0pt}{1.2ex}%
    \smash{\footnotesize second-best}%
  }%
  \endgroup
}

\newcolumntype{Y}{>{\centering\arraybackslash}X}
\usepackage{afterpage}

\newcommand{\agent}[1]{\textsc{#1}}

\usepackage{titletoc}

\title{\methodname{}: Adaptive \textbf{S}equential–\textbf{Par}allel Scaling with \textbf{C}ontext Management for Retrieval-Augmented Generation}

\author{
Yuxin Yang$^{1}$,
Gangda Deng$^{1}$,
{\"O}mer Faruk Akg{\"u}l$^{1}$,
Nima Chitsazan$^{2}$,
\textbf{Yash Govilkar$^{2}$},
\\
\textbf{Akasha Tigalappanavara$^{2}$,
Shi-Xiong Zhang$^{2}$,
Sambit Sahu$^{2}$,
Viktor Prasanna$^{1}$}
\\[0.5em]
$^{1}$University of Southern California \quad
$^{2}$Capital One
}

\begin{document}

\maketitle

\begin{abstract}

Retrieval-Augmented Generation (RAG) grounds large language model outputs in external evidence, but remains challenged on multi-hop question answering that requires long reasoning. Recent works scale RAG at inference time along two complementary dimensions: sequential depth for iterative refinement and parallel width for coverage expansion. However, naive scaling causes context contamination and scaling inefficiency, leading to diminishing or negative returns despite increased computation.
To address these limitations, we propose \methodname{},
a multi-agent framework that coordinates sequential and parallel inference-time scaling under a unified context management mechanism.
\methodname{} employs specialized agents that maintain a shared global context and provide explicit control over the scaling process. It generates targeted, complementary sub-queries for each branch to enable diverse parallel exploration, and explicitly regulates exiting decisions based on answer correctness and evidence grounding.
To optimize scaling behavior, we further introduce a lightweight fine-tuning method with process-level verifiable preferences, which improves the efficiency of sequential scaling and effectiveness of parallel scaling.
Across single- and multi-hop QA benchmarks, \methodname{} consistently outperforms previous RAG baselines, yielding an average +6.2 F1 improvement under lower inference cost.
\end{abstract}

\section{Introduction}\label{sec:intro}

Large Language Models (LLMs)~\citep{yang2025qwen3, achiam2023gpt} augmented with retrieval have become a dominant approach for knowledge-intensive tasks. Retrieval-Augmented Generation (RAG)~\citep{lewis2020rag} grounds LLM outputs in external evidence, providing access to current and factual information. Despite these advantages, standard RAG systems with a single retrieval and question answering step often struggle on complex queries, particularly multi-hop queries~\citep{gao2023retrieval,tang2024multihop} which demand deep reasoning and hypothesis exploration. To handle these intricate demands, recent works focus on scaling RAG at inference time along two dimensions: deepening the reasoning chain (sequential scaling or depth) ~\cite{trivedi2023ircot, wang2025deepnote, lee2024planrag} and broadening hypothesis exploration (parallel scaling or width)~\cite{wang2024speculativerag, li2024dmqrrag}. 

However, scaling is not straightforward: mechanically increasing depth and width results in redundant computation and thus \emph{Scaling Inefficiency}, while uncontrolled context growth introduces noise and leads to \emph{Context Contamination}; together, these effects cause performance to plateau and place the system on a suboptimal cost–accuracy trade-off curve~\citep{leng2024long, wang2025bee, lin2025refrag}.
More concretely, sequential scaling iteratively refines reasoning but suffers from context contamination as evidence accumulates indiscriminately, overwhelming the model's attention capacity and deteriorating performance~\citep{wu2025depth,misaki2025wider}. 
Parallel scaling increases coverage by exploring multiple reasoning paths, but it often produces redundant branches and relies on simple aggregation schemes (e.g., majority voting) that fail to synthesize complementary evidence~\citep{wang2025samplingefficient,xiao2025limopro}.

Although sequential depth and parallel width are intrinsically complementary — the former enables iterative refinement while the latter expands evidence coverage — this complementarity is difficult to realize.
Without explicit regulation of scaling behavior and information consolidation, their interaction amplifies redundancy and noise, undermining both efficiency and answer quality. This highlights the need for a framework that coordinates sequential refinement and parallel exploration under a unified context management mechanism.

Building on these insights, we 
propose \methodname{} (Adaptive \textbf{S}equential–\textbf{Par}allel Scaling with \textbf{C}ontext Management for Retrieval-Augmented Generation), a multi-agent framework that explicitly controls reasoning width, depth, and context during inference-time scaling.
Instead of indiscriminately increasing computation, \methodname{} organizes scaling around specialized collaborating agents. In particular, the scaling process is orchestrated by three key agents: a \agent{Query Rewriter} generates complementary sub-queries to maximize parallel diversity; an \agent{Answer Evaluator} determines when to continue or stop reasoning, preventing premature termination; and a \agent{Context Manager} consolidates evidence across sequential rounds and parallel branches, selectively integrating relevent information while filtering noise, thereby maintaining a compact and stable context. Together, these agents enable \methodname{} to achieve better accuracy-cost trade-offs than prior sequential-only or parallel-only approaches.


Furthermore, we design a lightweight fine-tuning strategy that leverages process-level preference pairs derived from scaling behaviors. These preference pairs guide the model toward more effective and efficient scaling by reducing redundant expansions and learning a more accurate stopping policy. We summarize our contributions as follows:
\begin{itemize}[
  leftmargin=0pt,
  labelwidth=0pt,
  labelsep=0pt,
  itemindent=0pt,
  align=parleft,
  itemsep=1.2pt,      
  topsep=2pt,       
  parsep=0pt  
]
    \item[\circnum{1}]\hspace{13pt}
 \textbf{Agent design principles for efficient scaling.}
    We identify \emph{Context Contamination} and \emph{Scaling Inefficiency} as the key challenges in inference-time RAG scaling and formulate agent-level design principles for mitigating them, including enhanced exploration diversity and context management across sequential depth and parallel width.
    \item[\circnum{2}]\hspace{13pt}
 \textbf{A collaborative multi-agent framework for efficient scaling.} 
    We introduce \methodname{}, a multi-agent sequential--parallel RAG scaling framework that controls the complementary width and depth scaling with a global information consolidation mechanism, generalizing prior sequential-only and parallel-only methods while achieving superior accuracy--cost trade-offs.
    \item[\circnum{3}]\hspace{13pt}
 \textbf{Scaling-oriented fine-tuning.} 
    We design a scaling-oriented fine-tuning strategy that supervises the scaling process. Preference pairs are constructed from verifiable intermediate metrics (evidence coverage and stopping accuracy), allowing the model to internalize when scaling is beneficial. This process-level calibration enhances both effectiveness and efficiency without heavy fine-tuning.
\end{itemize}

\section{Background and Related Work}
We review sequential scaling and parallel scaling in RAG, along with recent work that introduces limited adaptive control over these processes.

\paragraph{Sequential Scaling in RAG.}
Sequential RAG methods iteratively refine reasoning through multi-round retrieval and generation.
IRCoT~\citep{trivedi2023ircot} interleaves retrieval and chain-of-thought reasoning via iterative query reformulation.
Iter-RetGen~\citep{shao2023iterretgen} alternates retrieval and generation to gradually surface evidence.
Chain-of-RAG~\citep{wang2025corag} constructs stepwise retrieval chains with explicit supervision.
These methods demonstrate the value of deeper reasoning but typically \emph{rely on fixed iteration counts and accumulate evidence across steps}, which can lead to noisy context and deteriorated performance (context contamination)~\citep{wu2025depth,misaki2025wider}.

\paragraph{Parallel Scaling in RAG.}
Parallel approaches expand coverage by exploring multiple reasoning paths in parallel. DMQR-RAG~\citep{li2024dmqrrag} and MCTS-RAG~\citep{hu2025mcts} generate diverse query rewrites or reasoning branches to retrieve complementary evidence, while Speculative-RAG~\citep{wang2024speculativerag} drafts multiple candidate answers for verification. However, these methods typically \emph{use fixed branching factors and simple aggregation strategies} (e.g., majority voting), which limits cross-branch evidence synthesis~\citep{wang2025samplingefficient,xiao2025limopro}.

\paragraph{Adaptive and Planning-Based RAG.}
Recent work introduces adaptive behaviors and planning into RAG pipelines.
Self-RAG~\citep{asai2024selfrag} and following works~\citep{yan2024crag, xu2025simrag} evaluate retrieval quality to dynamically revise outputs. Planning-based approaches~\citep{lee2024planrag, verma2024plan, gu2025rapid} generate structured plans before retrieval, while question-decomposition methods~\citep{ammann2025questiondecomp} split complex queries into sub-questions and plans the following execution. DeepNote~\citep{wang2025deepnote} maintains a persistent note state across steps and employs a heuristic stopping rule.
However, these approaches do not treat sequential depth and parallel width as \emph{jointly controllable inference-time resources}, making scaling implicit and inefficient. This gap motivates a mechanism that coordinates sequential refinement, parallel exploration, and context evolution in a unified manner.

\begin{figure*}[t]
    \vspace{-20pt}
    \centering
    \includegraphics[width=0.95\linewidth]{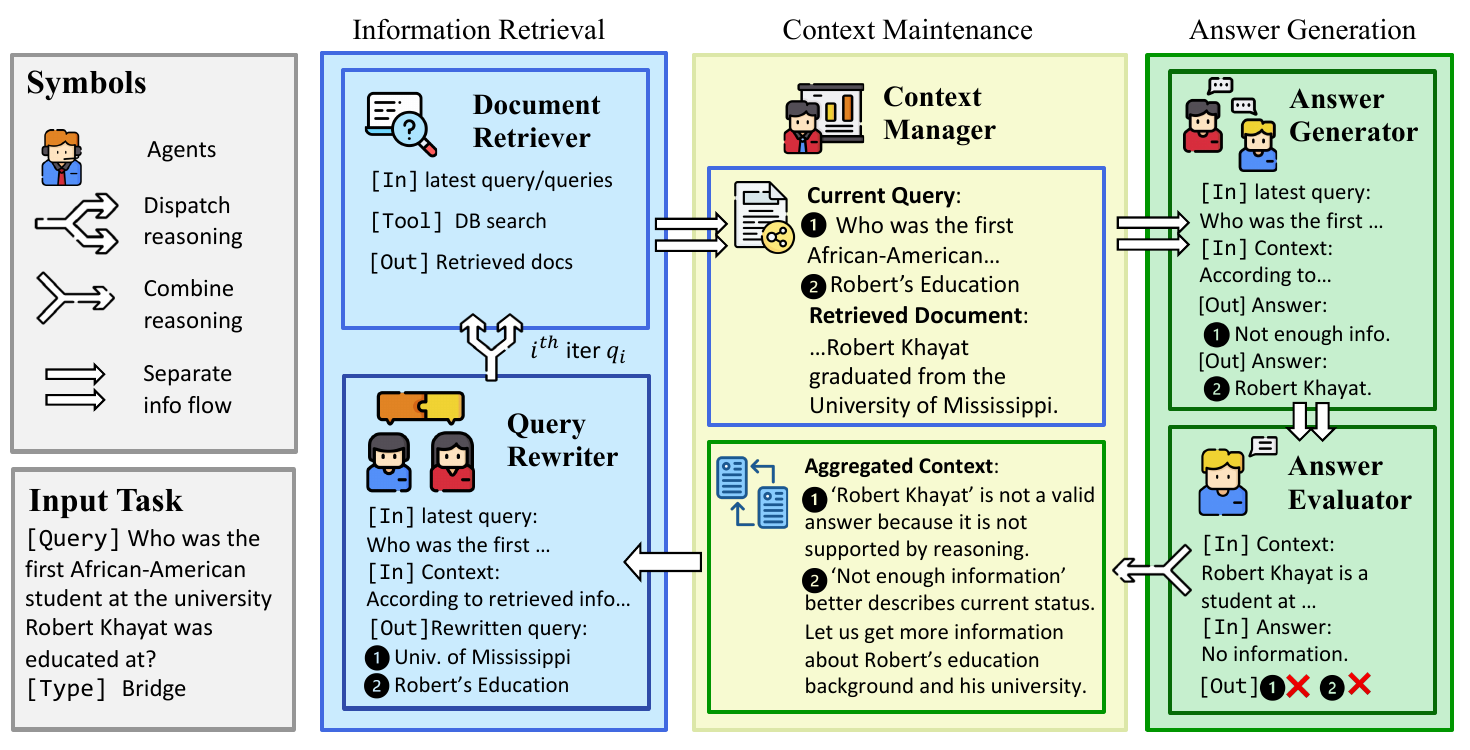}
    \vspace{-5pt}
    \caption{
        Overall \methodname{} framework.
        The figure illustrates the interactions, information flow, and decision signals across agents.
    }
    \vspace{-7pt}
    \label{fig:framework}
\end{figure*}
\section{Methods}\label{sec:method}

To address the \emph{Context Contamination} and \emph{Scaling Inefficiency} challenges identified in Section~\ref{sec:intro}, we introduce \methodname{}, a multi-agent framework that provides explicit control over depth, width, and information consolidation during inference-time scaling. We begin by formally characterizing scaling in RAG.

\subsection{Task Formulation \& System Overview}

\noindent\textbf{Problem Formulation.} We formulate the multi-hop QA task as a joint optimization process over Sequential Depth ($D$) and Parallel Width ($W$). Given an initial query $\boldsymbol{q}_0$ and an external corpus $\mathcal{C}$, \methodname{} models the reasoning process as a hierarchical composition of three core operators, denoted as $\Pi = S \circ P \circ I$ (formal formulation in Appendix~\ref{app:task_formulation}):

\begin{itemize}[
  leftmargin=0pt,
  labelwidth=0pt,
  labelsep=0pt,
  itemindent=0pt,
  align=parleft,
  itemsep=1.5pt,      
  topsep=4pt,       
  parsep=0pt   
]
    \item[(i)]\hspace{15pt}  \textbf{Single-path Inference Operator ($I$):} Executes a standard ``retrieve-reason-generate'' operation, as in naive RAG.
    \item[(ii)]\hspace{15pt} \textbf{Parallel Expansion Operator ($P$):} Dynamically generates $W^{(t)}$ parallel branches at the $t$-th inference step and aggregates information upon completion. 
    \item[(iii)]\hspace{15pt} \textbf{Sequential Refinement Operator ($S$):} Connects multiple parallel rounds in sequence, determining based on the current state whether to deepen the reasoning or terminate with the current answer. 
\end{itemize}

\begin{figure}[t]
    \centering
    \vspace{-7pt}
    \includegraphics[width=0.9\linewidth]{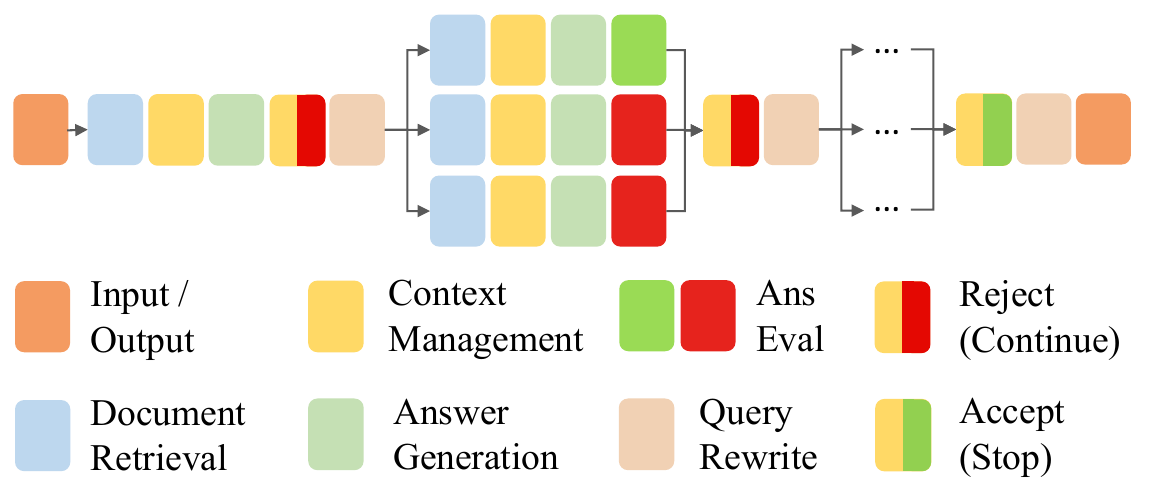}
    \caption{
        Single left-to-right rollout of the \methodname{} reasoning process. The Accept (Stop) and Reject (Continue) decisions are produced by the \agent{Context Manager} after merging the branches.
    }
    \vspace{-10pt}
    \label{fig:single-rollout}
\end{figure}

Moving from the abstract formulation to its system-level realization, we provide overview of system in Figure~\ref{fig:framework} and rollout in ~\ref{fig:single-rollout}. At round $t$, the \agent{Query Rewriter} generates parallel rewritten queries that form independent branches, where the \agent{Retriever} gathers evidence, the \agent{Context Manager} updates the shared context, and the \agent{Answer Generator} produces a candidate answer that is evaluated by the \agent{Answer Evaluator}. The \agent{Context Manager} then selects the best branch, merges evidence into a unified state, and decides whether to continue or stop reasoning based on the selected answer’s evaluation. Together with standard RAG components, the \agent{Context Manager}, \agent{Query Rewriter}, and \agent{Answer Evaluator} enable coordinated depth–width scaling with improved accuracy–cost trade-offs over prior sequential-only or parallel-only approaches.

\subsection{Agents} \label{agents}

\textbf{\agent{Retriever.}}
Given the current query $\boldsymbol{q}^{(t)}$, the retriever selects a small subset of passages from the corpus $\mathcal{C}$ using a similarity-based retrieval function followed by top-$k$ ranking.
$\sigma^{(t)}$ denotes the retrieval strategy used at this round.
\vspace{-4pt}
\[
r^{(t)} \sim \mathsf{Retrieve}_{\sigma^{(t)}}\!\big(\boldsymbol{q}^{(t)},\, C\big).
\]
\vspace{-4pt}

\noindent \textbf{\agent{Answer Generator.}}
Conditioned on the global query $\boldsymbol{q}_0$ and current memory $m^{(t)}$, the generator produces a candidate answer for this round using a language model.
\vspace{-4pt}
\[
a^{(t)} \sim \mathsf{AnsGenerate}_{\mathrm{LLM}}\!\big(\boldsymbol{q}_0,\, m^{(t)}\big).
\]
\vspace{-4pt}

\noindent \textbf{\agent{Answer Evaluator.}}
The evaluator evaluates the quality of the generated answer based on the query and current memory, generating a decision signal $e^{(t)}$ (e.g., \texttt{CONTINUE} or \texttt{STOP}).
\vspace{-4pt}
\[
\begin{aligned}
&e^{(t)} \ \sim
\mathsf{EvaluateAns}_{\mathrm{LLM}}\!\big(\boldsymbol{q}_0,\, \boldsymbol{q}^{(t)},\, a^{(t)},\, m^{(t)}\big).
\end{aligned}
\]
\vspace{-4pt}
Crucially, this prevents redundant iterations on simple queries, dynamically allocating the computational budget only to complex instances, reducing \emph{Scaling Inefficiency}.

\noindent \textbf{\agent{Context Manager.}}
This module lies in the center of the Context Management functionality. It maintains the evolving reasoning state by consolidating newly retrieved documents and generated candidate answers with the existing memory during scaling, mitigating \emph{Context Contamination}. Within each branch, it performs a query-aware update that attends to the current question and retains only the relevant evidence.
\vspace{-4pt}
\[
\begin{aligned}
m^{(t)} \sim \mathsf{MemUpdate}_{\mathrm{LLM}}\!\big(\boldsymbol{q}_0,\, \boldsymbol{q}^{(t)},\, m^{(t-1)},\, r^{(t)}, a^{(t)}\big).
\end{aligned}
\]
\vspace{-4pt}
This mechanism ensures the memory update is both \textbf{incremental} and \textbf{minimal}, thereby maintaining a consistent context window with low noise. 
After parallel branches complete, the \agent{Context Manager} orchestrates the aggregation of branches: it selects the most promising reasoning path based on answer quality and evidence support, then consolidates complementary information from all branches into a unified context.
\vspace{-4pt}
\[
\begin{aligned}
k^* &\leftarrow \mathsf{SelectBest}_{\mathrm{LLM}}\!\big(\{(a_k, m'_k)\}_{k=1}^{W}\big), \\
m^{(t)} &\leftarrow \mathsf{ContextMerge}_{\mathrm{LLM}}\!\big(m'_{k^*}, \{m'_k\}_{k \neq k^*}\big).
\end{aligned}
\]
\vspace{-4pt}
These operations alleviate \emph{Context Contamination} by maintaining a unified, query-focused memory state rather than concatenating all evidence without focus.

\noindent \textbf{\agent{Query Rewriter.}}
To operationalize the Parallel Expansion Operator ($P$) efficiently, this agent functions as a \emph{one-to-many generator}. Distinct from standard rewriters that sequentially refine a single query, it synthesizes $W^{(t)}$ intent-specific queries (and their corresponding retrieval strategies) in a \emph{single inference pass}. Conditioned on the global question $\boldsymbol{q}_0$, current memory $m^{(t)}$, and the target width $W^{(t)}$, it explores orthogonal retrieval directions to maximize information coverage, alleviating the \emph{Scaling Inefficiency}.
\vspace{-4pt}
\[ \begin{aligned} &\big(\{\sigma^{(t+1,k)}, q^{(t+1,k)}\}_{k=1}^{W^{(t)}}\big) \\&\hspace{1cm}\sim \mathsf{RewriteQuery}_{\mathrm{LLM}}\!\big(\boldsymbol{q}_0,\, \boldsymbol{q}^{(t)},\, m^{(t)}, W^{(t)}\big). \end{aligned} \]
\vspace{-4pt}

\subsection{Execution Protocol}
\label{sec:algorithm}
We formally describe the execution flow as a dynamic execution loop (summarized in Algorithm~\ref{alg:framework_flow}). The process begins with the user query $\boldsymbol{q}_0$. The memory state is initialized as empty, $m^{(0)} \leftarrow \emptyset$. \methodname{} regulates the consumption of computational resources ($\mathcal{B}$) through two mechanisms:

\paragraph{Parallel Distribution \& Branch Execution.}
To explore the search space (Width $W$), the system first triggers the \agent{Query Rewriter} to generate a set of intent-specific rewritten queries $\{q_k\}_{k=1}^W$.
Crucially, each rewritten query spawns an independent \textbf{parallel execution branch}. Within each branch $k$, the agents perform a full atomic update cycle:
\begin{itemize}[
  leftmargin=0pt,
  labelwidth=0pt,
  labelsep=0pt,
  itemindent=0pt,
  align=parleft,
  itemsep=1.5pt,      
  topsep=4pt,       
  parsep=0pt   
]
    \item[(i)]\hspace{15pt} Retrieve: The branch retrieves raw evidence $r_k$ relevant to the rewritten query.
    \item[(ii)]\hspace{15pt} Context Update: The \agent{Context Manager} compresses $r_k$ and integrates it with the previous history $m^{(t-1)}$ to form a branch-local memory  $m'_k$.
    \item[(iii)]\hspace{15pt} Generate \& Evaluate: Conditioned on updated state $m'_k$, answer $a_k$ is generated and evaluated. 
\end{itemize}
This phase yields $W$ fully processed hypothesis tuples $\{(m'_k, a_k, e_k)\}_{k=1}^W$.

\paragraph{Aggregation \& Global Control.}
Once the parallel branches complete, the system performs a select-and-decide operation:
\begin{itemize}[
  leftmargin=0pt,
  labelwidth=0pt,
  labelsep=0pt,
  itemindent=0pt,
  align=parleft,
  itemsep=1.2pt,      
  topsep=2pt,       
  parsep=0pt  
]
    \item[(i)]\hspace{15pt} Selection and Consolidation:
    The \agent{Context Manager} selects the most promising branch $k^*$ from the candidate paths $\{(a_k, m'_k)\}_{k=1}^W$ and consolidates evidence into a unified memory, preserving complementary information while reducing redundancy and noise.
    \item[(ii)]\hspace{15pt} Termination Decision: The system checks the decision signal of the selected winner ($e_{k^*}$). If $e_{k^*} == \texttt{STOP}$ or the budget $D_{max}$ is reached, the process terminates returning $a_{k^*}$. Otherwise, the loop continues to $t \leftarrow t+1$.
\end{itemize}

\newcommand{\RewriteQuery}{\mathsf{RewriteQuery}}
\newcommand{\Retrieve}{\mathsf{Retrieve}}
\newcommand{\MemUpdate}{\mathsf{MemUpdate}}
\newcommand{\AnsGenerate}{\mathsf{AnsGenerate}}
\newcommand{\EvaluateAns}{\mathsf{EvaluateAns}}
\newcommand{\SelectBest}{\mathsf{SelectBest}}
\newcommand{\ContextMerge}{\mathsf{ContextMerge}}
\begin{algorithm}[t]
\caption{\methodname{} Inference Algorithm}
\label{alg:framework_flow}
\begin{algorithmic}[1]
\Require Query $q_0$, Corpus $\mathcal{C}$, Max Depth $D_{\max}$
\State $m^{(0)} \gets \emptyset,\ t \gets 1$
\While{$t \le D_{\max}$}
  \State \textcolor{blue}{\textit{// Phase 1: Parallel Distribution}}
  \State $\{q_k\}_{k=1}^W \gets \RewriteQuery(q_0, m^{(t-1)})$

  \State \textcolor{blue}{\textit{// Phase 2: Sequential Execution}}
  \ForAll{$k \in \{1,\dots,W\}$ \textbf{in parallel}}
    \State $r_k \gets \Retrieve(q_k, \mathcal{C})$
    \State $m'_k \gets \MemUpdate(m^{(t-1)}, r_k)$
    \State $a_k \gets \AnsGenerate(q_0, m'_k)$
    \State $e_k \gets \EvaluateAns(a_k, m'_k)$
  \EndFor

  \State \textcolor{blue}{\textit{// Phase 3: Aggregation \& Control}}
  \State $k^* \gets \SelectBest_{\mathrm{LLM}}\big(\{(a_k, e_k, m'_k)\}_{k=1}^W\big)$
  \State $m^{(t)} \gets \ContextMerge(m'_{k^*}, \{m'_k\}_{k \neq k^*})$

  \If{$e_{k^*} = \texttt{STOP}$}
    \State \Return $a_{k^*}$
  \EndIf
  \State $t \gets t + 1$
\EndWhile
\end{algorithmic}
\end{algorithm}

\subsection{Scaling-oriented Fine-tuning}

To improve system performance without incurring high cost, we adopt a scaling-oriented, preference-based fine-tuning strategy. We construct preference pairs that reward higher combined paragraph recall for parallel scaling and more accurate stopping decisions for sequential scaling. Through these process-level preferences, the model is explicitly encouraged to achieve broader evidence coverage and more reliable stopping behavior, improving scaling effectiveness and efficiency.

\paragraph{Preference Data Collection for the Query Rewriter.}
The \agent{Query Rewriter} is supervised using paragraph-level retrieval recall. Rewrites that retrieve more gold-supporting evidence are preferred over those with lower recall.
Although this supervision signal is purely recall-based, it implicitly encourages \emph{diverse yet focused} sub-queries: improving recall typically requires accessing complementary evidence rather than duplicating the same retrieval trajectory. As a result, the trained rewriter reduces redundancy across parallel branches(Section~\ref{subsec:finetune}), making width expansion more effective.

\paragraph{Preference Data Collection for the Answer Evaluator.}
We construct preference data by sampling multiple \texttt{STOP}/\texttt{CONTINUE} decisions of \agent{Answer Evaluator} and assigning preferences based on the correctness of the corresponding answers.

A key challenge in supervising this decision is the asymmetry of error costs: a \emph{wrong stop} (accepting an incorrect answer, exiting prematurely) is typically far more damaging than a \emph{wrong continue} (rejecting a correct answer, continuing reasoning), as it eliminates the possibility of further corrections. 
To reflect this asymmetry, we introduce a weighting function $w(y^{+},y^{-})$ over preference pairs. For early-stop–critical cases in which the preferred action is \textsf{continue} but the rejected action is \textsf{stop}, we assign $w(y^{+},y^{-})=\lambda>1$; all other preference pairs retain unit weight. The weighted DPO loss for \agent{Answer Evaluator} is therefore
\vspace{-4pt}
\[
\begin{aligned}
\mathcal{L}_{\mathrm{AE}}(x,y^{+},y^{-})
&=
w(y^{+},y^{-})
\Big[
-\log \sigma\!\big(
\\[-2pt]
\beta[
&\log p_\theta(y^{+}\!\mid x)
-\log p_\theta(y^{-}\!\mid x)
]
\big)
\Big],
\end{aligned}
\]
\vspace{-4pt}
where $\lambda>1$ emphasizes penalties on incorrect early stopping while still allowing early termination on easy instances.

As we show in Section~\ref{subsec:finetune}, targeted fine-tuning calibrates stopping behaviour while encouraging diverse parallel exploration.




\section{Experiments}
\label{sec:experiments}

\subsection{Datasets and Metrics}
\begin{figure*}[ht]
    \centering
    \vspace{-12pt}
    \includegraphics[width=\textwidth]{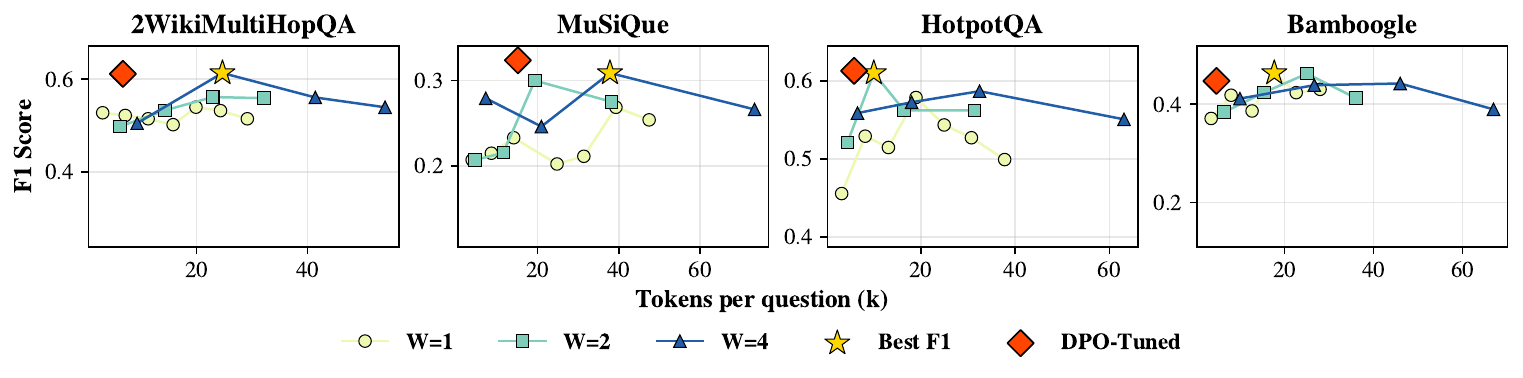}
    \vspace{-20pt}
    \caption{F1–cost tradeoff across $(W,D)$ configurations with Qwen2.5-7B-Instruct model. Each panel plots F1 versus tokens per question for one dataset. Curves correspond to widths $W{=}1,2,4$. For $W{=}1$, the sequential depths are $D{=}2,4,6,8,10,12,14$. For $W{=}2$ and $W{=}4$, the depths are $D{=}2,4,6,8$. Along each curve, points are ordered from left to right by increasing $D$. The gold star marks the best untuned $(W,D)$ configuration, and the red diamond denotes our fine-tuned model, which achieves comparable or better F1 at substantially lower token cost.}
    \label{fig:cost-vs-f1}
    \vspace{-6pt}
\end{figure*}

We evaluate \methodname{} on both single-hop and multi-hop QA benchmarks. 
For single-hop QA, we use \textbf{Natural Questions} (NQ; \citealt{kwiatkowski2019natural}).  
For multi-hop QA, we include 2-hop datasets \textbf{HotpotQA} \citep{yang2018hotpotqa}, \textbf{2WikiMultiHopQA} (or \textbf{2WikiMQA}) \citep{ho2020constructing}, 
\textbf{Bamboogle} \citep{press2023bamboogle}, and 2-4 hop dataset \textbf{MuSiQue} \citep{trivedi2022musique}.  
All training of the \agent{Query Rewriter} and \agent{Answer Evaluator} uses the training splits of datasets, and we report evaluation results on their test splits of 500 subsamples (detailed in Appendix \ref{app:datasets}).

We report Exact Match (EM), F1, and Accuracy following~\citet{vu2021ava} and ~\citet{asai2024selfrag}, and additionally use paragraph recall to assess retrieval effectiveness.
Cost is measured by total token usage (input + output) and latency.


\subsection{Backbone Models}
\label{subsec:backbone}

All agent components are instantiated using
Qwen2.5-7B-Instruct~\citep{qwen2025qwen25} and Qwen3-32B~\citep{yang2025qwen3}.
These two backbones enable us to evaluate \methodname{} under different levels of reasoning capacity.

\subsection{Retrievers}
\label{subsec:retrievers}

We employ a hybrid retrieval stack for \methodname{} across all experiments, as is the common practice in RAG systems ~\citep{thakur2021beir, seo2019real, zhao2024retrieval}. \methodname{} uses an agentic selection of sparse and dense retrieval (see Appendix~\ref{app:retrieval_methods}) by default. As most baselines only support using a single retrieval method, ablation results of using only one retrieval method are reported in Appendix~\ref{app:ablation_retrieval}, demonstrating that using one retrieval method does not deteriorate performance and keeping the fairness of comparison with other baseline methods.

\subsection{Baselines}

We compare \methodname{} against strong inference-time scaling paradigms in RAG:
(i) sequential refinement methods IRCoT~\citep{trivedi2023ircot},
Iter-RetGen~\citep{shao2023iterretgen}, and
DeepNote~\citep{wang2025deepnote};
(ii) sequential self-reflective correction method Self-RAG~\citep{asai2024selfrag};
and (iii) parallel drafting--verification method Speculative-RAG~\citep{wang2024speculativerag}. See Appendix~\ref{app:baseline_configs} for more details.

\subsection{Implementation Details}

All experiments are conducted on a multi-GPU machine equipped with four NVIDIA RTX~PRO~6000 Blackwell Server Edition GPUs (96GB each) and dual Intel Xeon 6730P processors (128 logical cores). LLMs are served using \texttt{vLLM}~(v0.10.2) with continuous batching and tensor parallelism. 

For all results, the sequential controller is capped at a maximum depth of~8. We use fixed $W=2$ for all reported results. For \methodname{} without fine-tuning, we use $D=6$ for Qwen-2.5-7B-Instruct and $D=8$ for Qwen3-32B to avoid confounding performance with noisy termination signals.

We finetune the base LLM using the combined training data from single-step generation of \agent{Query Rewriter} and \agent{Answer Evaluator}, trained with LoRA~\citep{hu2022lora}; additional hyperparameters are provided in Appendix~\ref{app:implementation_details}.

\section{Results}\label{sec:results}

\begin{table*}[t]
\vspace{-20pt}
\tiny
\setlength{\tabcolsep}{2.4pt}
\renewcommand{\arraystretch}{0.96}
\centering
\caption{End-to-end performance across QA benchmarks. \bestlegend{} and \secondlegend{} results are highlighted. For some datasets, we report DeepNote\textsuperscript{*}results using the numbers provided in the original paper.}
\vspace{-5pt}
\label{tab:main-results}
\begin{tabularx}{\textwidth}{l *{20}{N}}
\toprule
\multirow{3}{*}{\textbf{Methods \& LLMs}}
  & \multicolumn{4}{c}{\textbf{Single-hop}}
  & \multicolumn{16}{c}{\textbf{Multi-hop}} \\
\cmidrule(lr){2-5}\cmidrule(lr){6-21}
  & \multicolumn{4}{c}{\textbf{NQ}}
  & \multicolumn{4}{c}{\textbf{HotpotQA}}
  & \multicolumn{4}{c}{\textbf{2WikiMQA}}
  & \multicolumn{4}{c}{\textbf{MusiQue}}
  & \multicolumn{4}{c}{\textbf{Bamboogle}} \\
\cmidrule(lr){2-5}\cmidrule(lr){6-9}\cmidrule(lr){10-13}\cmidrule(lr){14-17}\cmidrule(lr){18-21}
  & {acc.} & {f1} & {em} & {avg.}
  & {acc.} & {f1} & {em} & {avg.}
  & {acc.} & {f1} & {em} & {avg.}
  & {acc.} & {f1} & {em} & {avg.}
  & {acc.} & {f1} & {em} & {avg.} \\
\midrule
\rowcolor{gray!20}
\multicolumn{21}{l}{\textit{Vanilla RAG}}\\
Qwen2.5-7B-Instruct
    & 26.2 & 13.3 & 2.4 & 14.0
    & 42.4 & 42.5 & 32.0 & 39.0
    & 35.6 & 38.5 & 32.6 & 35.6
    & 11.4 & 13.8 & 6.6 & 10.6
    & 14.4 & 17.7 & 12.0 & 14.7 \\
Qwen3-32B
    & 28.2 & 16.0 & 4.2 & 16.1
    & 51.2 & 37.3 & 21.8 & 36.8
    & 48.4 & 32.2 & 14.2 & 31.6
    & 15.8 & 10.6 & 2.4 & 9.6
    & 25.6 & 17.2 & 6.4 & 16.4 \\
\midrule                                                                   
\rowcolor{gray!20}
\multicolumn{21}{l}{\textit{Baselines using Qwen2.5-7B-Instruct}}\\
Parallel (Naive)
& 27.6 & 13.0 & 3.0 & 14.5                                      
& 41.0 & 42.4 & 32.6 & 38.7                                               
& 35.6 & 38.2 & 32.0 & 35.3                                                
& 11.0 & 13.8 & 6.4 & 10.4                                                 
& 19.2 & 24.0 & 15.2 & 19.5 \\         
Sequential (Naive)
&28.6&13.6&3.2&15.1
&43.2&41.3&30.2&38.2
&35.6&37.1&30.0&34.2
&10.8&12.1&5.2&9.4
&19.2&22.5&15.2&19.0\\
IRCoT
  & 39.6 & 17.3 & 4.6 & 20.5
    & 40.4 & 33.8 & 23.2 & 32.5
    & 40.0 & 33.2 & 23.6 & 32.3
    & 11.8 & 12.4 & 4.8 & 9.7
    & 19.2 & 16.7 & 12.0 & 16.0 \\

Iter-RetGen
  & \secondbest{55.2} & 46.5 & 35.2 & 45.6
    & 48.0 & 50.2 & 38.6 & 45.6
    & 43.4 & 37.9 & 29.4 & 36.9
    & 19.8 & 21.4 & 13.0 & 18.1
    & 32.0 & 36.6 & 28.0 & 32.2 \\

Speculative-RAG
  & 40.6 & 41.5 & 31.4 & 37.8
  & 29.2 & 34.6 & 28.2 & 30.7
  & 39.6 & 41.6 & 36.8 & 39.3
  & 9.8 & 15.7 & 8.0 & 11.2
  & 16.8 & 19.8 & 16.0 & 17.5 \\

Self-RAG
  & 37.6 & 37.2 & 26.6 & 33.8
  & 42.6 & 46.0 & 35.4 & 41.3
  & 40.8 & 41.3 & 35.2 & 39.1
  & 11.6 & 15.8 & 8.0 & 11.8
  & \secondbest{37.6} & 41.1 & 29.6 & 36.1 \\
  
DeepNote
  & \best{52.8} & \best{53.3} & \best{40.2} & \best{48.8}
  & \secondbest{50.6}\textsuperscript{*} & \secondbest{59.2}\textsuperscript{*} & \secondbest{48.0}\textsuperscript{*} & \secondbest{52.6}\textsuperscript{*}
  & 50.0\textsuperscript{*} & 51.4\textsuperscript{*} & 41.8\textsuperscript{*} & 47.7\textsuperscript{*}
  & 14.6\textsuperscript{*} & 19.8\textsuperscript{*} & 11.6\textsuperscript{*} & 15.3\textsuperscript{*}
  & 26.4 & 36.3 & 24.8 & 29.2 \\

\textbf{\methodname{} (Ours)}
  & 45.4 & 46.9 & 35.6 & 42.6
  & 48.4 & 55.9 & 45.4 & 49.9 
  & \secondbest{56.2} & \secondbest{56.2} & \secondbest{45.2} & \secondbest{52.5}
  & \secondbest{17.8} & \secondbest{24.3} & \secondbest{16.2} & \secondbest{19.4}
  & 36.8 & \secondbest{42.7} & \secondbest{36.0} & \secondbest{38.5} \\

\textbf{\methodname{} + DPO (Ours)}
  & \secondbest{51.4} & \secondbest{51.8} & \secondbest{40.0} & \secondbest{47.7}
  & \best{53.8} & \best{61.3} & \best{49.4} & \best{54.8}
  & \best{63.0} & \best{61.2} & \best{48.4} & \best{57.5}
  & \best{25.2} & \best{32.4} & \best{20.8} & \best{26.1}
  & \best{38.4} & \best{44.7} & \best{33.6} & \best{38.9} \\

\midrule
\rowcolor{gray!20}
\multicolumn{21}{l}{\textit{Baselines using Qwen3-32B}}\\
Parallel (Naive)
&31.2&16.4&4.4&17.3
&51.0&38.1&23.6&37.6
&44.6&32.9&16.6&31.4
&14.8&10.9&2.6&9.4
&27.2&16.6&6.4&16.7\\
Sequential (Naive)
&31.4&15.6&4.0&17.0
&51.2&37.4&22.2&36.9
&45.2&31.5&14.8&30.5
&14.6&10.3&1.8&8.9
&28.8&15.4&4.0&16.1\\
IRCoT
  & 33.2 & 22.7 & 11.2 & 22.4
    & 34.4 & 36.1 & 27.2 & 32.6
    & 24.4 & 19.6 & 12.0 & 18.7
    & 6.6  & 10.1 & 5.4 & 7.4
    & 25.6 & 12.9 & 7.2 & 15.2 \\
Iter-RetGen
  & 54.4 & 47.4 & 34.6 & 45.5
    & 52.4 & 61.4 & 48.6 & 54.1
    & 53.8 & 54.7 & 42.2 & 50.2
    & 23.2 & 29.4 & 18.6 & 23.7
    & 40.8 & 47.3 & 33.6 & 40.6 \\
Speculative-RAG
  & 45.6 & 48.0 & 36.0 & 43.2
  & 33.6 & 39.0 & 31.2 & 34.6
  & 31.2 & 32.7 & 29.4 & 31.1
  & 13.8 & 19.1 & 10.8 & 14.6
  & 32.0 & 37.3 & 31.2 & 33.5 \\
Self-RAG
  & 45.4 & 44.9 & 31.8 & 40.7
  & 53.0 & 57.9 & 45.0 & 52.0
  & 55.8 & 55.3 & 46.2 & 52.4
  & 12.0 & 20.7 & 10.4 & 14.4
  & 49.6 & 54.6 & 42.4 & 48.9 \\
DeepNote
    & \best{58.4} & 41.3 & 26.4 & 42.0
    & \secondbest{60.2} & \secondbest{65.2} & \secondbest{51.2} & \secondbest{58.9}
    & 69.2 & 64.4 & 50.6 & 61.4
    & \secondbest{30.8} & 30.9 & 18.6 & 26.8
    & 40.8 & 47.0 & 33.6 & 40.5 \\
\textbf{\methodname{} (Ours)}
  & 53.6 & \best{54.7} & \best{39.4} & \best{49.2}
    & \best{62.6} & \best{68.7} & \best{55.0} & \best{62.1}
    & \best{78.4} & \best{74.1} & \best{61.2} & \best{71.2}
    & \best{31.8} & \best{36.6} & \best{23.8} & \best{30.7}
    & \best{56.0} & \best{63.0} & \best{49.6} & \best{56.2} \\
\textbf{\textbf{\methodname{}} + DPO  (Ours)}
  & \secondbest{56.6} & \secondbest{51.1} & \secondbest{38.0} & \secondbest{48.6}
    & 58.2 & 65.0 & 51.0 & 58.1
    & \secondbest{75.2} & \secondbest{69.8} & \secondbest{57.6} & \secondbest{67.5}
    & 28.8 & \secondbest{33.9} & \secondbest{21.4} & \secondbest{28.0}
    & \secondbest{53.6} & \secondbest{59.5} & \secondbest{44.8} & \secondbest{52.6} \\
\bottomrule
\vspace{-5pt}
\end{tabularx}
\end{table*}

\begin{figure}[t]
  \centering
    \vspace{-12pt}
  
  \includegraphics[width=0.95\linewidth]{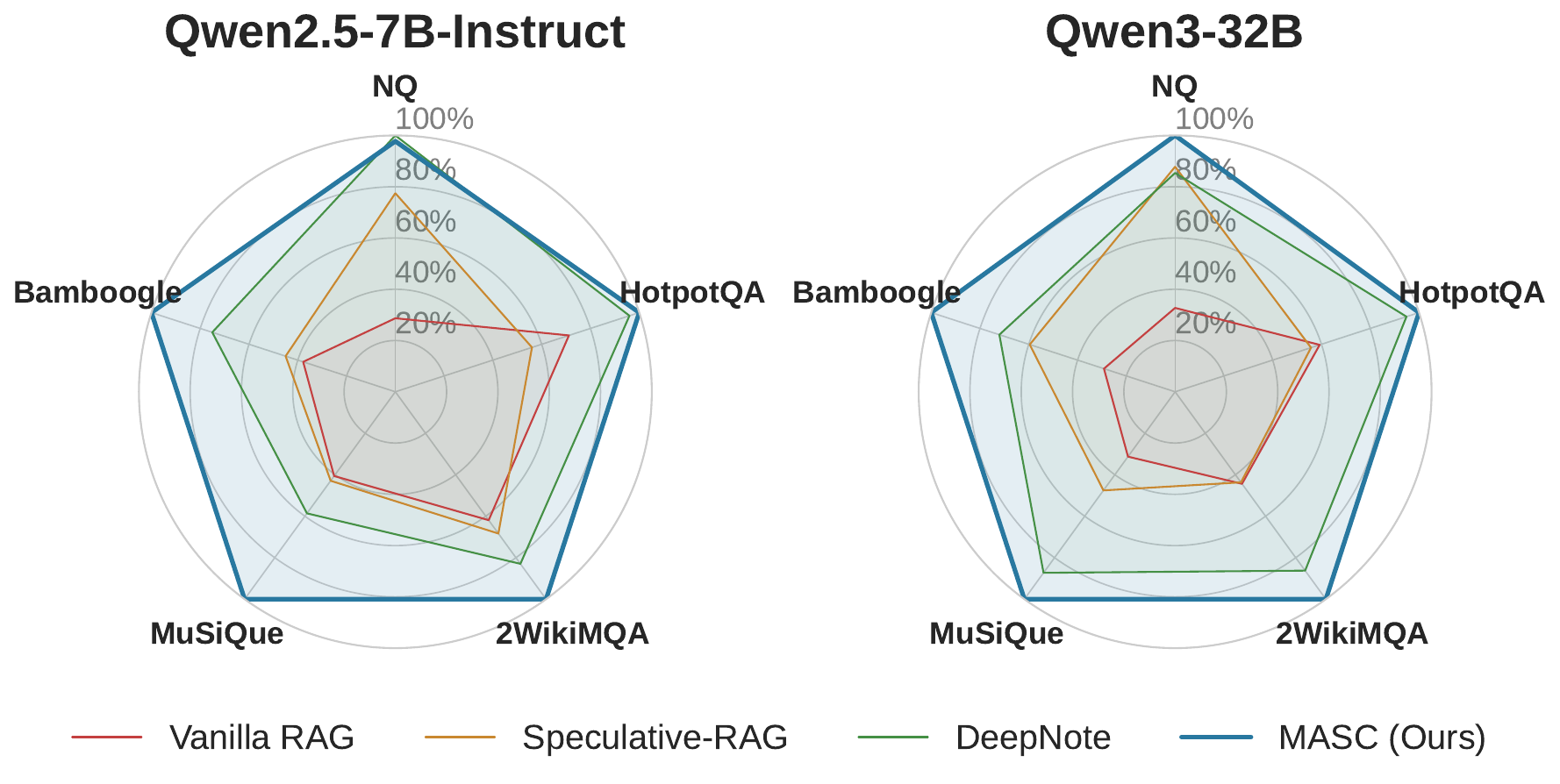}
  \caption{
  Relative performance of different RAG systems using two backbone LLMs across five datasets.
  Each plot compares Vanilla RAG, the representative parallel (Speculative-RAG) and sequential (DeepNote) RAG scaling methods, and \methodname{} (ours).
  }
  \label{fig:radar-normalized}
    \vspace{-10pt}
\end{figure}

Figure~\ref{fig:radar-normalized} and
Table~\ref{tab:main-results} summarize end-to-end performance across QA datasets. 

\textbf{Without fine-tuning}, \methodname{} with joint depth-width scaling consistently matches or exceeds strong sequential- or parallel-only baselines. \methodname{} yields substantial gains, especially on multi-hop QA datasets. On MuSiQue (2-4 hops) with Qwen2.5-7B-Instruct model for example, \methodname{} achieves 24.3 F1 score compared to 19.8 of DeepNote (previous SOTA), demonstrating the benefit of unified depth-width control on complex queries even with base models.

\textbf{After fine-tuning}, \methodname{} achieves matched or higher accuracy with significantly lower cost. \methodname{}+DPO with Qwen2.5-7B-Instruct backbone improves average F1 score on multihop QA datasets by 5.1 and reduces the cost by \textasciitilde 60\%. For Qwen3-32B,  \methodname{}+DPO slightly decreases the performance but reduces the average token cost by \textasciitilde 90\% (Figure~\ref{fig:cost-vs-f1}, details in Table~\ref{tab:dpo-ablation}). Compared with the previous best RAG scaling baseline, \methodname{} achieves +6.19 F1 score with 52.2\% less token cost on average (see Appendix~\ref{app:token-deepnote}). 

These results validate our two main claims: (i) joint sequential-parallel scaling substantially strengthens multi-hop reasoning compared to existing single-direction scaling, and (ii) targeted lightweight fine-tuning improves both answer quality and computational efficiency.

\subsection{Analysis of Sequential Depth and Parallel Width}
\label{sec:rq1_analysis}

To better understand the gains in joint scaling, we first analyze how sequential depth and parallel width individually and jointly contribute to the performance and cost of \methodname{}.



\begin{figure}[t]
    \centering
    \vspace{-5pt}
    \includegraphics[width=\linewidth]{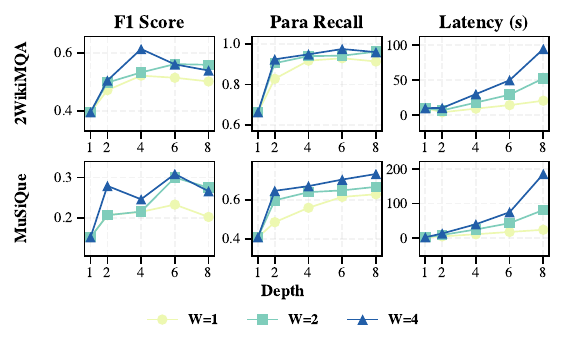}
    \vspace{-20pt}
    \caption{Answer quality (F1 scores, top row) and retrieval performance (paragraph recall, bottom row) on 2WikiMQA and 
    MuSiQue for $W{=}1,2,4$ using Qwen2.5-7B-Instruct model.}
    \label{fig:acc_retrieval_vs_depth}
    \vspace{-10pt}
\end{figure}

\paragraph{Depth and width provide complementary benefits under scaling.}
Figure~\ref{fig:acc_retrieval_vs_depth} shows accuracy curves for 2WikiMQA and MuSiQue under $W{=}1,2,4$. Increasing depth yields gains from $D{=}1$ to $D{=}4$ overall. Beyond this point, quality plateaus or even slightly decreases. 
On the other hand, increasing width improves answer quality when depth-only scaling saturates. Across both datasets, the paragraph recall grows consistently with width, indicating that parallel branching explores more diverse reformulations of the question and retrieval of more evidence, and directly translates into increased F1 score. Latency increases approximately linearly with depth and width, implying a rapidly rising cost at larger scaling levels. This motivates a closer examination of how different scaling choices affect the cost–quality tradeoff.


\paragraph{Cost–quality tradeoffs.}
Figure~\ref{fig:cost-vs-f1} provides a direct comparison of performance at matched compute budgets. 
Across all configurations, $W{=}2$ strikes the best balance between answer quality and cost. It outperforms $W{=}1$ configuration by introducing diversity in reasoning and in retrieval, while avoiding the sharp cost increases associated with $W{=}4$. We observe that the best performance–cost trade-off emerges at moderate depth and width rather than extreme scaling in either direction consistently on all datasets.
For the results reported in Table~\ref{tab:main-results}, we uniformly employed $W{=}2$ based on this observation. Moreover, the fact that the optimal depth varies across datasets justifies the adaptive, query-dependent depth controller that adjusts the reasoning length according to the difficulty of each query.

Finally, fine-tuning (Section~\ref{subsec:finetune}) yields a more favorable F1–cost operating point: it achieves matching or higher F1 scores to the $W\times D$ grid search at substantially lower token costs by diversifying parallel exploration (\agent{Query Rewriter}) and avoiding wasteful continuation (\agent{Answer Evaluator}). For example, on 2WikiMultiHopQA, the fine-tuned model at $W{=}2$ matches the base model's best performance across all $W\times D$ configurations while using only \textasciitilde 30\% tokens.

\subsection{Fine-Tuning for Improved Sequential and Parallel Scaling} \label{subsec:finetune}

In this subsection, we provide an in-depth analysis of how targeted fine-tuning of the base LLM enhances the scaling behavior of \methodname{}. 


\paragraph{Larger document coverage in Query Rewriting.}


Fine-tuning directly improves the recall of retrieval by improving the quality of rewritten queries. We conduct a controlled comparison on samples where the tuned model proceeds to the second retrieval iteration (first iteration for rewriting) and show that fine-tuning consistently improves retrieval recall across all datasets. For 7B model, recall improvements range from +2.3\% (MuSiQue) to +36.1\% (NQ), with an average gain of +13.1\% (Appendix~\ref{app:finetune-recall}). The 32B model exhibits a smaller gain of +0.5\% to +5.2\% across datasets.

To further assess how finetuning affects the structure of query rewrites, we measure the average pairwise Jaccard overlap between the document sets retrieved with rewritten queries, following the definition in Appendix~\ref{app:def-jaccard}. 
Fine-tuning reduces the overlap among retrieved documents by up to 27.4\%(NQ), indicating that rewritten queries become less redundant and more focused. This reduction emerges naturally from optimizing for higher recall, as improved recall requires retrieving complementary rather than overlapping evidence.

\begin{figure}[t]
    \centering
    \vspace{-20pt}
    \includegraphics[width=1.0\columnwidth]{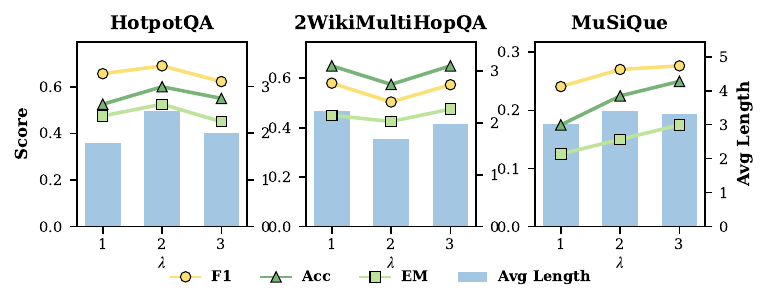}
    \vspace{-8pt}
    \caption{Effect of the weighting parameter $\lambda$ on \agent{Answer Evaluator} behavior evaluated on a 100-sample test subset.}
    \label{fig:ae_lambda_dataset}
    \vspace{-14pt}
\end{figure}

\paragraph{Weighted DPO leads to more conservative stopping and deeper reasoning.}
Figure~\ref{fig:ae_lambda_dataset} shows per-dataset trends as we vary the weighting parameter $\lambda$, which up-weights \emph{wrong-stop} errors during training. Increasing $\lambda$ strengthens the penalty on premature termination, leading the fine-tuned evaluator to adopt a more conservative stopping criterion and reducing early exits on difficult instances.

On more complex MuSiQue (2–4 hops), this results in deeper rollouts and higher F1, as additional reasoning enables multi-step evidence aggregation. On 2-hop 2WikiMQA dataset, however, large $\lambda$ can slightly degrade performance due to unnecessary continuation and occasional distractors. We use $\lambda{=}2$ in our final model, as it provides a favorable trade-off across datasets—improving robustness on harder benchmarks while avoiding over-conservative behavior on simpler ones.

\begin{figure}[t]
    \centering
    \vspace{-14pt}
    \includegraphics[width=0.8\columnwidth]{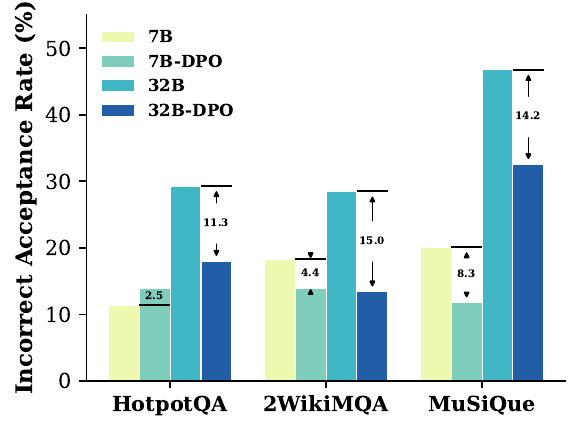}
    \caption{Incorrect acceptance rates across datasets for base and DPO-finetuned models.}
    \vspace{-10pt}
    \label{fig:incorrect_accept}
\end{figure}

\paragraph{Reduced incorrect acceptance rates during Answer Evaluation.}
Fine-tuning substantially reduces the tendency to accept incorrect intermediate answers.  
As shown in Figure~\ref{fig:incorrect_accept}, larger base models (32B) exhibit higher over-acceptance rates than the 7B model, reflecting stronger overconfidence, while DPO fine-tuning consistently mitigates this failure—reducing incorrect acceptance by $11$--$15$ points on 32B and $3$--$8$ points on 7B across datasets. These behaviors translate into higher end-to-end QA performance (Table~\ref{tab:main-results}).

\section{Conclusions}
We introduced \methodname{}, a multi-agent framework that addresses \emph{Context Contamination} and \emph{Scaling Inefficiency} in RAG scaling. 
Through specialized control agents that jointly regulate scaling along {sequential depth} and {parallel width}, \methodname{} coordinates how computation is expanded across branches and rounds while consolidating complementary evidence into a unified context state, yielding substantial gains across single-hop and multi-hop QA benchmarks.
We further introduced a scaling-targeted, preference-based fine-tuning strategy that encourages complementary parallel queries and improves the stability of stopping decisions, thereby reducing unnecessary expansions and achieving better cost–performance characteristics for \methodname{}.

\section{Limitations}
Our approach focuses on improving retrieval-augmented generation through agentic test-time scaling and structured control over depth and width of reasoning. As a result, the proposed method introduces additional inference-time overhead compared to standard single-pass RAG pipelines, which may limit its applicability in latency-critical settings. While our design aims to make this overhead cost-aware, the trade-off between performance and efficiency depends on the quality of the stopping and aggregation mechanisms.
Moreover, our experiments are conducted on established open-domain question answering benchmarks with retrieval preferences. Performance may vary under different retrieval systems, corpus characteristics, or domain-specific knowledge bases. Finally, while the proposed framework improves robustness to incomplete retrieval by exploring multiple reasoning paths, it does not guarantee correctness when retrieved evidence is systematically misleading or missing.

\section{License and Ethics}
This work uses publicly available datasets and pretrained language models, and does not involve the collection of personal or sensitive data. The proposed method does not introduce risks beyond those already associated with large language models and retrieval-augmented generation systems.
As with other RAG-based approaches, generated outputs may reflect biases or inaccuracies present in the underlying models or databases. We emphasize that the method is intended for research purposes and should be applied with appropriate safeguards when used in real-world or high-stakes scenarios.

Parts of this paper were refined using AI-assisted writing tools to improve clarity and presentation, and minor code edits were conducted with AI coding assistance under human supervision. All technical ideas, experimental design choices, and final implementations were created and verified by the authors, and the AI tools were used only as auxiliary aids rather than sources of substantive intellectual contributions.
\clearpage

\bibliography{ref}

\AtEndDocument{
  \clearpage
\appendix
\startcontents[app]
\printcontents[app]{l}{1}{%
  \section*{Appendix}%
}

\section{Experiment details}\label{app:experiment_details}

\subsection{Datasets}
\label{app:datasets}

We evaluate our method on five knowledge-intensive QA benchmarks,
covering both single-hop and multi-hop reasoning.

\paragraph{Natural Questions (NQ).}
Natural Questions \citep{kwiatkowski2019natural} is a single-hop open-domain QA benchmark consisting of real anonymized Google search queries. In our experiments, we follow the AdaptiveRAG evaluation setup ~\citep{shen2024adaptiverag}, which introduces a 500-query test subset. The retrieval corpus is the standard DPR Wikipedia passage collection ~\citep{karpukhin2020dense}, chunked into 100-word passages.
Indexing follows the dense retrieval pipeline described in the AdaptiveRAG documentation.

\paragraph{HotpotQA.}
HotpotQA \citep{yang2018hotpotqa} is a 2-hop QA benchmark
with supporting-fact supervision.
Following IRCoT \citep{trivedi2023ircot},
we use the 500-query subsampled test set provided by the IRCoT repository.\footnote{\url{https://github.com/StonyBrookNLP/ircot}}
The corpus is constructed using the IRCoT Wikipedia passage-building procedure.

\paragraph{2WikiMultiHopQA.}
2WikiMultiHopQA \citep{ho2020constructing} contains 2-hop questions requiring cross-entity bridging inference based on Wikipedia pages.
We again follow IRCoT \citep{trivedi2023ircot} and use the 500-query subsampled test set distributed in the IRCoT processed-data package.
The corpus is the same IRCoT Wikipedia passage corpus.

\paragraph{MuSiQue.}
MuSiQue \citep{trivedi2022musique} is a challenging 2--4 hop QA benchmark created by compositional reasoning over multiple Wikipedia paragraphs.
As with the previous datasets, we use the 500-query subsampled test set following IRCoT \citep{trivedi2023ircot}, and apply the same corpus construction procedure as IRCoT.

\paragraph{Bamboogle.}
Bamboogle \citep{press2023bamboogle} is a 2-hop QA dataset targeting adversarial or low-resource settings, with a small official test set. 
We use the official test split from the HuggingFace distribution.\footnote{\url{https://huggingface.co/datasets/chiayewken/bamboogle}}
Following EvoRAG \citep{zhang2025evorag}, we evaluate Bamboogle using a unified Wikipedia passage corpus.

\subsection{Indexing}
\label{app:indexing}

We build separate retrieval indices for each dataset corpus. The indexing
pipeline consists of the following components:

\paragraph{Document Chunking.}
In the retrieval corpus, documents are
organized at the paragraph level, where each paragraph is associated with a title and treated as an independent retrieval unit. For datasets that use Wikipedia as the corpus (i.e., Natural Questions, Bamboogle), we use the DPR
Wikipedia passages\footnote{\url{https://dl.fbaipublicfiles.com/dpr/wikipedia_split/psgs_w100.tsv.gz}}, which splits Wikipedia articles into 100-word passages.

\paragraph{Sparse Indexing.}
We use Elasticsearch 7.10.2 to build BM25 indices. Each document is indexed with the following fields: \texttt{title}, \texttt{paragraph\_text},
\texttt{paragraph\_index}, \texttt{url}, and \texttt{is\_abstract}. The \texttt{title} and \texttt{paragraph\_text} fields are analyzed using the
English analyzer for stemming and stopword removal.

\paragraph{Dense Indexing.}
We use BGE-large-en-v1.5 (\texttt{BAAI/bge-large-en-v1.5} hosted on Huggingface) as the embedding model to encode all documents into dense vectors. The embeddings are indexed using FAISS~\citep{faiss} with a flat inner product (FlatIP) index for exact nearest
neighbor search. At query time, we compute the dot product similarity between the query embedding and all document embeddings.

\subsection{Retrieval methods}
\label{app:retrieval_methods}

Our system supports two retrieval methods, which can be dynamically selected
by the language model during inference:

\paragraph{Sparse Retrieval (BM25).}
Given a query, we perform BM25 retrieval against the Elasticsearch index. The query is preprocessed by removing wh-words (e.g., ``what'', ``where'', ``when'') to improve lexical matching. We retrieve the top-$k$ documents ranked by BM25 score, where $k=6$ for all baselines and our method.

\paragraph{Dense Retrieval.}
Given a query, we encode it using the same BGE-large-en-v1.5 model used for indexing and retrieve the top-$k$ documents based on dot product similarity with the document embeddings. We use the same $k=6$ as sparse retrieval.

\paragraph{Dynamic Retrieval Selection.}
Unlike prior methods that rely on a single fixed retrieval approach, our agentic system treats different retrieval methods as \emph{tools} that can be intelligently selected based on the query characteristics. At each retrieval
step, the language model decides which retrieval method to invoke. This design choice is motivated by the complementary strengths of sparse and dense retrieval: BM25 excels at lexical matching and handling rare entities, while
dense retrieval better captures semantic similarity.

We acknowledge that this introduces a potential fairness consideration when comparing against baselines, as most prior methods use a single retrieval method. However, we note that: (i) these baseline methods do not natively support dynamic retrieval selection in their architectures, (ii) there is limited motivation for them to use multiple retrieval methods since they lack the decision-making mechanism to choose between them, and (iii) ablation study for \methodname{} using only dense retrieval in Appendix ~\ref{app:ablation_retrieval} indicates that the improvement does not originate from the dynamic retrieval selection.

\paragraph{Note on \texttt{global\_max\_num\_paras}.}
For all baseline methods that accumulate retrieved passages into an ever-growing context, we set a unified
\texttt{global\_max\_num\_paras} threshold. Without such a cap, these methods may receive excessively long contexts across iterations, which is known to harm LLM
generation quality and distort fair comparison. This design follows the IRCoT~\citep{trivedi2023ircot}
implementation, where limiting the accumulated context is necessary to prevent the model from being overwhelmed by large, redundant evidence pools.

In contrast, note-based or context-compression methods, i.e., Deepnote~\citep{wang2025deepnote} and our \methodname, do not rely on raw accumulation of retrieved passages. Their internal representations remain compact, so \texttt{global\_max\_num\_paras} is not applicable to them.

\subsection{Implementation Details} \label{app:implementation_details}

\paragraph{Hardware.}
All training and inference workloads run on a single server with
four NVIDIA RTX~PRO~6000 Blackwell Server Edition GPUs
(97{,}887~MiB each; 384~GB total), dual Intel Xeon~6730P CPUs
(2~sockets × 32~cores × 2~threads = 128 logical cores), 
and 1.0~TiB RAM.

\paragraph{Software Environment.}
Experiments are executed under Python~3.11.13 with the following key
libraries: PyTorch~2.8.0{+}cu128, vLLM~0.10.2, Transformers~4.56.1,
PEFT~0.17.1, TRL~0.23.1, and DeepSpeed~0.18.0. Sparse retrieval is
implemented using Elasticsearch~7.10.2. Dense retrieval does not rely
on FAISS in our deployment and is served via model-backed embedding
queries.

\paragraph{Inference Configuration.}
Generation modules use temperature~0.5, top-$p=1.0$, and a maximum generation length of 600 tokens. Answer Evaluator modules operate deterministically with temperature~0. Parallel branches use global random seed 42. The sequential controller is allowed up to 8 reasoning rounds per query.

\paragraph{Training Hyperparameters.}
The DPO training use LoRA adapters with rank~16, 
$\alpha{=}32$, dropout 0.05, and target modules
\texttt{q\_proj}, \texttt{k\_proj}, \texttt{v\_proj}, \texttt{o\_proj}.
DPO training uses learning rate 5e--5 (LoRA),
5e--7 (full fine-tuning), batch size 1 with gradient accumulation~4 
(effective batch size~16 across 4~GPUs), maximum sequence length 8192,
BF16 precision, and DPO $\beta{=}0.1$.  We use the AdamW optimizer and train for 3 epochs.

\paragraph{Retriever Configuration.}
Sparse retrieval uses Elasticsearch 7.10.2 with BM25 scoring. Dense
retrieval uses model-encoded embeddings served via lightweight
FastAPI endpoints. The same retrieval stack is used for all baselines
to ensure fair comparison.

\subsection{Baseline Configurations}
\label{app:baseline_configs}

We compare against the following representative RAG methods, categorized by their scaling strategy:

\paragraph{Vanilla RAG} is a single-step retrieval baseline: given a query, the system retrieves the top-$k$ documents using dense retrieval and generates an answer conditioned on the concatenation of the query and retrieved passages. No iterative refinement or parallel exploration is performed.

\paragraph{Parallel (Naive)} is a naive parallel scaling baseline that generates $W$ independent answers at temperature 0.7, then aggregates via majority voting (self-consistency). This baseline uses the same answer generation prompt as Vanilla RAG but samples multiple responses to leverage diversity. The final answer is selected by counting exact-match votes among the postprocessed candidates. We sweep $W \in \{2, 4, 6, 8, 10\}$ and report the best result for each dataset.

\paragraph{Sequential (Naive)} is a naive sequential scaling baseline that implements a verify-and-retry strategy. The system generates an answer, then prompts the same LLM to self-evaluate and output a confidence score (0--1). If confidence falls below a threshold (0.7), the system retries up to $D{=}3$ attempts. The final answer is selected from the attempt with highest self-evaluated confidence.

\paragraph{IRCoT}~\citep{trivedi2023ircot} is an interleaved retrieval method with Chain-of-Thought prompting. The model alternates between generating a reasoning step and retrieving relevant documents, iterating until the answer is produced. We use the official implementation\footnote{\url{https://github.com/StonyBrookNLP/ircot}} and follow its original design, which utilizes a BM25 sparse retriever, as the method relies heavily on keyword-level lexical overlap.

\paragraph{Iter-RetGen}~\citep{shao2023iterretgen} is an iterative retrieval-generation method that alternates between retrieval and generation, using the previous generation as context for the next retrieval query. Since no official implementation is publicly available, we re-implemented this method following the algorithm described in the original paper.

\paragraph{Speculative-RAG}~\citep{wang2024speculativerag} is a parallel drafting approach that generates multiple draft answers using different retrieved document subsets and selects the best response via a verification step. Since no official implementation is publicly available, we re-implemented this method based on the original paper. We use $k{=}6$ document subsets with $m{=}10$ drafts for 2WikiMultiHopQA and MuSiQue, and $k{=}3$ subsets with $m{=}5$ drafts for the other datasets, with a maximum of 15 retrieved documents.

\paragraph{Self-RAG}~\citep{asai2024selfrag} is a self-reflective retrieval-augmented generation method that learns to retrieve, generate, and critique its own output through special reflection tokens. Rather than using the original fine-tuned model, we adopt the LangGraph-based re-implementation\footnote{\url{https://langchain-ai.github.io/langgraph/tutorials/rag/langgraph_self_rag/}} which reproduces the Self-RAG workflow using standard LLMs without specialized fine-tuning. This enables fair comparison using the same backbone models (Qwen2.5-7B-Instruct and Qwen3-32B) across all methods.

\paragraph{DeepNote}~\citep{wang2025deepnote} is a sequential RAG method that maintains a structured note representation across iterations, compressing retrieved evidence into a running summary. We use the official implementation\footnote{\url{https://github.com/thunlp/DeepNote/tree/main}} but replace the retrieval backend and generator LLM to match our unified evaluation setup.

\section{Extended Definitions} \label{app:def}

\subsection{Task Formulation} \label{app:task_formulation}

We formalize the retrieval-augmented reasoning task that our system aims to solve.
The objective is to answer a complex query $q_0$ by iteratively interacting with an external corpus $C$ and refining intermediate hypotheses over multiple rounds of reasoning.
To facilitate a precise formulation, we introduce the notations used throughout this section in Table ~\ref{tab:notations}.

\begin{table}[h]
\centering
\caption{Notation summary for the RAG task formulation.}
\begin{tabular}{p{0.11\textwidth} p{0.32\textwidth}}
\toprule
\textbf{Symbol} & \textbf{Description} \\
\midrule
$C$ & Corpus available for retrieval. \\
$q_0$ & Global (initial) query. \\
$q^{(t)}$ & Query at round $t$. \\
$m^{(t)} \in \mathcal{M}$ & Memory state at round $t$. \\
$a^{(t)} \in \mathcal{A}$ & Candidate answer at round $t$. \\
$D$ & Total number of reasoning rounds (depth). \\
$W^{(t)}$ & Number of parallel reasoning branches at round $t$ (width). \\
\bottomrule \label{tab:notations}
\end{tabular}
\end{table}

We describe the reasoning process as a hierarchical composition of three operators:
an \emph{intra-path} operator $\mathsf{I}$, a \emph{parallel} operator $\mathsf{P}$, and a
\emph{sequential} operator $\mathsf{S}$. The overall system can be expressed as
\[
\Pi(q_{0}, C) \triangleq (\mathsf{S} \circ \mathsf{P} \circ \mathsf{I})(q_{0}, C).
\]

The operator $\mathsf{I}$ performs one round of retrieval-augmented inference along a single reasoning path, updating its memory and optionally producing a candidate answer. Note that an \emph{operator} in our formulation is an abstract computational role, rather than a concrete agent.

The parallel operator $\mathsf{P}^{(t)}$ applies $\mathsf{I}$ across $W^{(t)}$ branches at round $t$ and aggregates their outcomes:
\[
\begin{aligned}
&\mathsf{P}^{(t)}(q^{(t)}, m^{(t-1)}, C) \\
&=
\mathsf{Merge}\!\Big(
    \big\{
        \mathsf{I}(q^{(t,k)}, m^{(t-1,k)}, C)
    \big\}_{k=1}^{W^{(t)}}
\Big).
\end{aligned}
\]

Finally, the sequential operator composes the round-specific parallel processes:
\[
\mathsf{S}\ \!\big(\{ \mathsf{P}^{(t)} \}_{t=1}^{D}\big)(q_{0}, C)
=
\big(\mathsf{P}^{(D)} \circ \cdots \circ \mathsf{P}^{(1)}\big)(q_{0}, C),
\]
where $D$ denotes the number of reasoning rounds until termination.

This formulation captures the hierarchical structure of multi-step RAG: sequential composition models refinement over time, while parallel composition captures the breadth of exploration within each round.

\subsection{Jaccard Overlap of Query Groups.} \label{app:def-jaccard}
Given a set of rewritten queries generated for the same question, we measure the degree of redundancy between them using the Jaccard overlap of their retrieved document sets.  
For a rewritten query $q$, let $\mathcal{R}(q)$ denote the set of top-$k$ retrieved paragraphs obtained from the dense retriever. For two rewritten queries $q_i$ and $q_j$, their Jaccard overlap is defined as
\[
\mathrm{Jac}(q_i, q_j)
=
\frac{\,|\mathcal{R}(q_i) \cap \mathcal{R}(q_j)|\,}
     {\,|\mathcal{R}(q_i) \cup \mathcal{R}(q_j)|\,}.
\]
This quantity ranges from $0$ (no shared retrieved evidence) to $1$ (identical retrieval results).

For a group of rewrites $\{q_1,\ldots,q_m\}$, we report the 
\emph{average pairwise Jaccard overlap}, defined as
\[
\mathrm{JacAvg}
=
\frac{2}{m(m-1)}
\sum_{i < j}
\mathrm{Jac}(q_i, q_j).
\]
Lower overlap indicates that the rewrites retrieve more complementary evidence, whereas high overlap suggests that rewrites collapse to similar retrieval trajectories.  
In our analysis, we compute this metric separately for DPO-chosen and DPO-rejected rewrites; the former consistently exhibit lower overlap, indicating that recall-based supervision implicitly encourages more diverse and evidence-complementary sub-queries.

\section{Additional Results}
\subsection{Token cost before and after fine-tuning}
\begin{table}[ht]
  \centering
  \tabcolsep=3pt
  \caption{DPO fine-tuning result: accuracy and efficiency comparison between fixed-depth baseline and DPO-tuned model with adaptive depth control. 7B uses $D{=}6$, $W{=}2$; 32B uses $D{=}8$, $W{=}2$.}
  \label{tab:dpo-ablation}
  \small
  \resizebox{\linewidth}{!}{
  \begin{tabular}{@{}llcccccc@{}}
  \toprule
  \multirow{2}{*}{\textbf{Model}} & \multirow{2}{*}{\textbf{Dataset}} & \multicolumn{2}{c}{\textbf{F1 (\%)}} &
  \multicolumn{2}{c}{\textbf{Avg Depth}} & \multicolumn{2}{c}{\textbf{Tokens (K)}} \\
  \cmidrule(lr){3-4} \cmidrule(lr){5-6} \cmidrule(lr){7-8}
  & & Fixed & Tuned ($\Delta$) & Fixed & Tuned & Fixed & Tuned \\
  \midrule
  \multirow{5}{*}{7B}
   & HotpotQA    & 55.9 & 61.3 \posdelta{(+5.4)}  & 6 & 1.8 & 20.2 & 6.5 \\
   & MuSiQue     & 24.3 & 32.4 \posdelta{(+8.1)}  & 6 & 3.0 & 23.6 & 16.5 \\
   & 2WikiMQA    & 56.2 & 61.2 \posdelta{(+5.0)}  & 6 & 2.0 & 22.0 & 7.8 \\
   & NQ          & 34.7 & 51.8 \posdelta{(+17.1)} & 6 & 1.7 & 16.2 & 7.1 \\
   & Bamboogle   & 42.7 & 44.7 \posdelta{(+2.0)}  & 6 & 1.6 & 24.3 & 5.3 \\
  \midrule
  \multirow{5}{*}{32B}
   & HotpotQA    & 68.7 & 65.0 \negdelta{(-3.7)}  & 8 & 1.4 & 39.0 & 4.6 \\
   & MuSiQue     & 36.6 & 33.9 \negdelta{(-2.7)}  & 8 & 1.7 & 53.2 & 7.2 \\
   & 2WikiMQA    & 74.1 & 69.8 \negdelta{(-4.3)}  & 8 & 1.8 & 88.1 & 7.0 \\
   & NQ          & 54.7 & 51.1 \negdelta{(-3.6)}  & 8 & 1.1 & 129.6 & 2.9 \\
   & Bamboogle   & 63.0 & 57.9 \negdelta{(-5.1)}  & 8 & 1.3 & 104.6 & 3.9 \\
  \bottomrule
  \end{tabular}
  }
\end{table}

Table~\ref{tab:dpo-ablation} presents the impact of DPO fine-tuning on answer evaluation accuracy and computational efficiency. For the 7B model, DPO training yields consistent improvements across all five datasets, with F1 gains ranging from +2.0\% (Bamboogle) to +17.1\% (NQ). Notably, these accuracy improvements are accompanied by substantial efficiency gains: the tuned model reduces average retrieval depth from 6 to 1.6--3.0 iterations, resulting in 56--78\% reduction in token consumption.

For the 32B model, while DPO tuning did not improve the answer performance, it achieves dramatic efficiency improvements---reducing depth from 8 to 1.1--1.8 iterations and cutting token usage by 87--98\%. We attribute this to a distribution mismatch: the DPO training data was generated using the 7B model, causing the 32B model to adopt overly conservative early-stopping behavior. Despite this, the 32B tuned model still achieves competitive absolute performance (e.g., 69.8\% F1 on 2WikiMQA) while consuming only 7--8\% of the baseline's computational budget.

These results suggest that DPO fine-tuning effectively teaches the model to recognize answer sufficiency, enabling adaptive depth control that balances accuracy and efficiency. Future work could explore model-specific DPO training to better preserve the capabilities of larger models.

\subsection{Comparison of token cost with other methods} \label{app:token-deepnote}
Table~\ref{tab:token-cost-comparison} compares the computational cost of our DPO-tuned models against DeepNote, a state-of-the-art adaptive RAG baseline that employs iterative retrieval with early stopping. We did not report the cost comparison with other methods as their performance are significantly worse than these two methods. We conduct the comparison using both 7B and 32B model variants to ensure a fair comparison.

For the 32B model, our approach achieves substantial token savings across all five datasets, with an average reduction of 64.0\% compared to DeepNote. 
For the 7B model, on single-hop datasets (NQ and Bamboogle), our method reduces token consumption by 15.3\% and 28.2\% respectively and achieves a 14.1\% average reduction while maintaining competitive accuracy.

A key advantage of our approach lies in learning \emph{when} to stop retrieval through direct preference optimization. While DeepNote uses heuristic stopping criteria, our method learns a fine-grained stopping policy from preference data, leading to more efficient scaling behaiviors.

\begin{table}[t]
  \centering
  \tabcolsep=4pt
  \caption{Token cost comparison between DeepNote and our DPO-tuned models. Values represent average tokens per query. As we directly report the results from the DeepNote paper for some datasets using the Qwen2.5-7B-Instruct model, token costs on some datasets are marked missing '-'.}
  \label{tab:token-cost-comparison}
  \small
  \resizebox{\linewidth}{!}{
  \begin{tabular}{@{}lrrrrrr@{}}
  \toprule
  \multirow{2}{*}{\textbf{Dataset}} & \multicolumn{3}{c}{\textbf{7B Model}} & \multicolumn{3}{c}{\textbf{32B Model}} \\
  \cmidrule(lr){2-4} \cmidrule(lr){5-7}
  & DeepNote & Ours & $\Delta$ & DeepNote & Ours & $\Delta$ \\
  \midrule
  HotpotQA   & 6,235  & 6,511  & \posdelta{+4.4\%}  & 10,561 & 4,640 & \negdelta{-56.1\%} \\
  MuSiQue    & --     & 16,477 & --                  & 13,206 & 7,166 & \negdelta{-45.7\%} \\
  2WikiMQA   & --     & 7,785  & --                  & 11,416 & 6,981 & \negdelta{-38.8\%} \\
  NQ         & 8,359  & 7,081  & \negdelta{-15.3\%} & 18,816 & 2,935 & \negdelta{-84.4\%} \\
  Bamboogle  & 7,452  & 5,348  & \negdelta{-28.2\%} & 17,102 & 3,887 & \negdelta{-77.3\%} \\
  \midrule
  \textbf{Avg} & \textbf{7,349} & \textbf{6,313} & \textbf{\negdelta{-14.1\%}} & \textbf{14,220} & \textbf{5,122} & \textbf{\negdelta{-64.0\%}} \\
  \bottomrule
  \end{tabular}
  }
\end{table}



\subsection{Analysis of DPO fine-tuning on paragraph recall} 
To isolate the effect of DPO fine-tuning on query rewriting quality, we conduct a controlled comparison on samples where the tuned model proceeds to the second retrieval iteration (Table~\ref{tab:recall-comparison}).

The results reveal that DPO fine-tuning consistently improves recall across all datasets. For the 7B model, there is an average recall improvement of +13.1\%. The substantial improvement on NQ (+36.1\%) is particularly notable: the base model achieves only 18.7\% recall on these samples, indicating poor query rewriting for single-hop questions, while the tuned model reaches 54.8\%. For multi-hop datasets, improvements are more modest but consistent (+2.3\% to +7.3\%), suggesting that DPO training enhances the model's ability to generate effective follow-up queries.

The 32B model exhibits similar trends with gains of +0.5\% to +5.2\% across datasets. The smaller improvement magnitude compared to 7B can be attributed to two factors: (1) fewer samples reach the second iteration (e.g., only 29 NQ samples for 32B vs.\ 157 for 7B), limiting statistical power, and (2) the base 32B model already produces higher-quality rewrites, leaving less room for improvement.

These findings demonstrate that DPO fine-tuning improves \emph{how} queries are rewritten when additional retrieval is necessary.
\label{app:finetune-recall}
\begin{table}[t]
  \centering
  \tabcolsep=4pt
  \caption{Recall comparison on samples where the DPO-tuned model proceeds to the second retrieval iteration. This controlled comparison isolates the effect of DPO on query rewriting quality, excluding samples that terminate early.}
  \label{tab:recall-comparison}
  \small
  \begin{tabular}{@{}llcccc@{}}
  \toprule
  \textbf{Model} & \textbf{Dataset} & \textbf{Samples} & \textbf{Base} & \textbf{Tuned} & \textbf{$\Delta$} \\
  \midrule
  \multirow{4}{*}{7B}
   & HotpotQA  & 209 & 70.6 & 77.3 & \posdelta{+6.7} \\
   & MuSiQue   & 348 & 58.8 & 61.1 & \posdelta{+2.3} \\
   & 2WikiMQA  & 271 & 77.6 & 84.9 & \posdelta{+7.3} \\
   & NQ        & 157 & 18.7 & 54.8 & \posdelta{+36.1} \\
  \midrule
  \multirow{4}{*}{32B}
   & HotpotQA  & 104 & 67.8 & 71.6 & \posdelta{+3.8} \\
   & MuSiQue   & 224 & 55.5 & 60.0 & \posdelta{+4.5} \\
   & 2WikiMQA  & 246 & 81.3 & 86.5 & \posdelta{+5.2} \\
   & NQ        &  29 & 39.1 & 39.7 & \posdelta{+0.5} \\
  \bottomrule
  \end{tabular}
\end{table}

\begin{table}[ht]
\centering
\caption{Ablation study comparing adaptive retrieval with dense-only retrieval, using finetuned model Qwen-2.5-7B-Instruct-DPO. The grey texts shows differences compared to hybrid retrieval.}
\label{tab:ablation_dense_only}
\resizebox{\linewidth}{!}{
\begin{tabular}{lccc|ccc}
\toprule
\multirow{2}{*}{Dataset} 
& \multicolumn{3}{c|}{Hybrid Retrieval} 
& \multicolumn{3}{c}{Dense-Only} \\
& EM & F1 & Acc 
& EM & F1 & Acc \\
\midrule
HotpotQA 
& 49.4 & 61.3 & 53.8 
& 48.8 {\scriptsize\color{gray}($-$0.6)} 
& 62.1 {\scriptsize\color{gray}(+0.8)} 
& 53.0 {\scriptsize\color{gray}($-$0.8)} \\

MuSiQue 
& 20.8 & 32.4 & 25.2
& 20.4 {\scriptsize\color{gray}($-$0.4)}
& 31.7 {\scriptsize\color{gray}($-$0.7)}
& 25.4 {\scriptsize\color{gray}(+0.2)} \\

2WikiMQA
& 48.4 & 61.2 & 63.0
& 49.2 {\scriptsize\color{gray}(+0.8)}
& 61.6 {\scriptsize\color{gray}(+0.4)}
& 62.8 {\scriptsize\color{gray}($-$0.2)} \\

NQ 
& 40.0 & 51.8 & 51.4
& 39.2 {\scriptsize\color{gray}($-$0.8)}
& 52.5 {\scriptsize\color{gray}(+0.7)}
& 52.6 {\scriptsize\color{gray}(+1.2)} \\

Bamboogle 
& 33.6 & 44.7 & 38.4
& 33.6 {\scriptsize\color{gray}(0.0)}
& 44.5 {\scriptsize\color{gray}($-$0.2)}
& 40.0 {\scriptsize\color{gray}(+1.6)} \\
\midrule

Average 
& 38.4 & 50.3 & 46.4
& 38.2 {\scriptsize\color{gray}($-$0.2)}
& 50.5 {\scriptsize\color{gray}(+0.2)}
& 46.8 {\scriptsize\color{gray}(+0.4)} \\
\bottomrule
\end{tabular}
}
\end{table}
\subsection{Ablation on retrieval methods}\label{app:ablation_retrieval}
Most baselines in our comparison only use a single retrieval method. To enable a fair comparison, we therefore evaluate our framework under a \emph{dense-only} setting, matching the retrieval configuration used by the baselines. Although mixed retrieval is a common and robust practice in RAG systems, prior work~\cite{thakur2021beir} has also noted that strong dense retrievers often perform comparably to hybrid configurations when the corpus is well-structured and the embedding model is high quality. In this ablation, the retriever is fixed to use dense retrieval for all queries.

Table~\ref{tab:ablation_dense_only} reports results across five datasets. Overall, the differences between the adaptive hybrid setting and dense-only retrieval are small and inconsistent across datasets. 

These results suggest that dense retrieval alone is sufficiently effective for our task. This finding indicates that the primary performance improvements of our system stem from other components, such as the iterative query rewriting and answer evaluation modules, rather than the retrieval strategy selection.

\subsection{Case study}
To illustrate why increasing parallel width improves multi-hop retrieval, we present a representative HotpotQA example (Figure~\ref{fig:case-study}). The system is asked: \textit{``Where is Anticimex’s parent company headquartered?''}  
Under a single path (W=1), the system fails to retrieve the document about EQT—the parent company—making the answer unreachable. With two parallel branches (W=2), different sub-queries explore complementary directions: one identifies EQT as the parent company, while the other retrieves EQT’s headquarters. Together they recover all necessary evidence and produce the correct final answer.

\begin{figure*}[t]
    \centering
    \includegraphics[width=\textwidth]{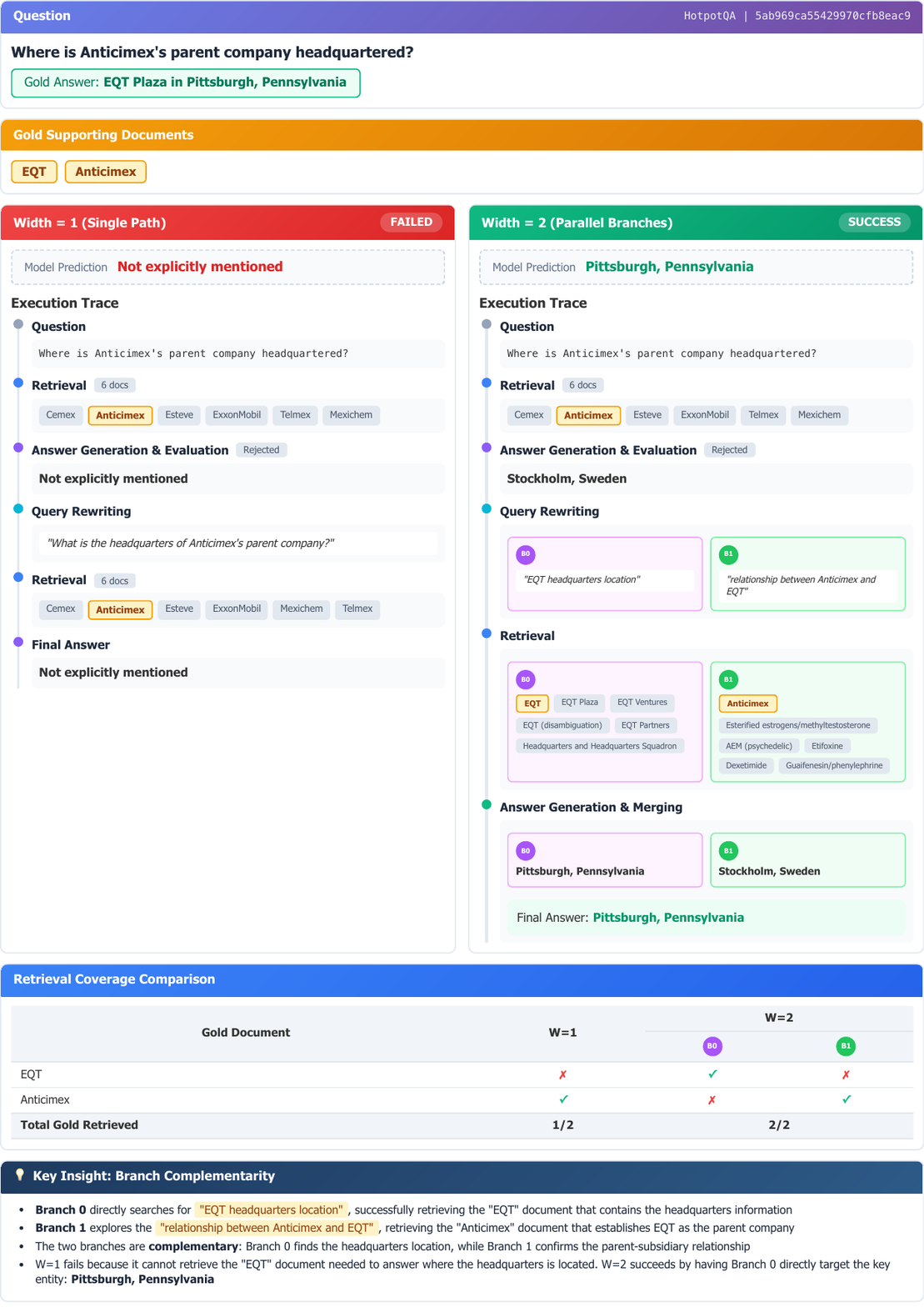}
    \caption{Case study illustrating how W=2 retrieves complementary evidence that W=1 misses.}
    \label{fig:case-study}
\end{figure*}

\paragraph{Analysis.}
The example in Figure~\ref{fig:case-study} reveals a clear pattern:

\begin{itemize}
    \item \textbf{W=1 fails due to insufficient retrieval.}  
          The single branch retrieves ``Anticimex'' but never surfaces ``EQT'', preventing the system from answering the headquarters question.

    \item \textbf{W=2 enables specialization.}  
          One rewritten query targets the parent-company relationship; the other explicitly seeks EQT’s headquarters. This division of labor increases coverage.

    \item \textbf{Complementarity resolves ambiguity.}  
          Only when the two branches are combined does the system access all gold evidence (both Anticimex and EQT documents).

    \item \textbf{Merging selects the supported answer.}  
          Although branches propose different answers, the system resolves them by grounding final selection in retrieved evidence.
\end{itemize}

This case concretely demonstrates our central claim:  
\textbf{parallel width expands the search space and recovers evidence that sequential refinement alone cannot reach}.

\onecolumn
\subsection{Width $\times$ Depth Grid Search Results}
\label{app:grid-search}
\begin{table*}[htbp]
  \scriptsize
  \setlength{\tabcolsep}{3.5pt}
  \renewcommand{\arraystretch}{1.1}
  \centering
  \caption{Width $\times$ Depth grid search results for
  Qwen2.5-7B-Instruct on multi-hop QA datasets.}
  \label{tab:width-depth-results}
  \begin{tabularx}{\textwidth}{l *{20}{N}}
  \toprule
  \multirow{2}{*}{\textbf{W$\times$D}}
    & \multicolumn{5}{c}{\textbf{HotpotQA}}
    & \multicolumn{5}{c}{\textbf{2WikiMQA}}
    & \multicolumn{5}{c}{\textbf{MuSiQue}}
    & \multicolumn{5}{c}{\textbf{Bamboogle}} \\
  \cmidrule(lr){2-6}
\cmidrule(lr){7-11}
\cmidrule(lr){12-16}
\cmidrule(lr){17-21}

    & {acc.} & {f1} & {em} & {avg.} & {tok.}
    & {acc.} & {f1} & {em} & {avg.} & {tok.}
    & {acc.} & {f1} & {em} & {avg.} & {tok.}
    & {acc.} & {f1} & {em} & {avg.} & {tok.} \\
  \midrule
  1$\times$2 & 37.0 & 45.5 & 35.0 & 39.2 & 3.2k & 57.0 & 52.7 &
  44.0 & 51.2 & 3.1k & 11.0 & 20.7 & 10.0 & 13.9 & 4.2k & 30.4 &
  37.1 & 29.6 & 32.4 & 3.5k \\
  1$\times$4 & 44.0 & 52.9 & 38.0 & 45.0 & 8.2k & 56.0 & 52.2 &
  42.0 & 50.1 & 7.1k & 15.0 & 21.5 & 13.0 & 16.5 & 10.6k & 36.0 &
  41.8 & 34.4 & 37.4 & 8.0k \\
  1$\times$6 & 41.0 & 51.4 & 38.0 & 43.5 & 13.2k & 52.0 & 51.4 &
  44.0 & 49.1 & 11.3k & 18.0 & 23.3 & 16.0 & 19.1 & 17.3k & 31.2 &
  38.6 & 30.4 & 33.4 & 12.7k \\
  1$\times$8 & 46.0 & 57.9 & 45.0 & 49.6 & 19.0k & 50.0 & 50.2 &
  42.0 & 47.4 & 15.8k & 15.0 & 20.2 & 13.0 & 16.1 & 24.8k & 39.2 &
  46.3 & 36.8 & 40.8 & 17.7k \\
  2$\times$2 & 43.0 & 52.1 & 42.0 & 45.7 & 3.8k & 53.0 & 49.8 &
  42.0 & 48.3 & 5.5k & 15.0 & 20.7 & 12.0 & 15.9 & 4.1k & 33.6 &
  38.4 & 32.0 & 34.7 & 5.8k \\
  2$\times$4 & 50.0 & 61.0 & 46.0 & 52.3 & 9.4k & 56.0 & 53.3 &
  43.0 & 50.8 & 13.6k & 12.0 & 21.6 & 11.0 & 14.8 & 11.1k & 35.2 &
  42.4 & 35.2 & 37.6 & 14.6k \\
  2$\times$6 & 50.0 & 56.2 & 42.0 & 49.4 & 15.8k & 61.0 & 56.1 &
  44.0 & 53.7 & 22.2k & 26.0 & 30.0 & 21.0 & 25.7 & 18.8k & 36.8 &
  46.2 & 34.4 & 39.1 & 24.3k \\
  2$\times$8 & 46.0 & 56.2 & 41.0 & 47.7 & 30.7k & 62.0 & 55.9 &
  42.0 & 53.3 & 31.6k & 21.0 & 27.5 & 18.0 & 22.2 & 37.7k & 36.8 &
  41.3 & 33.6 & 37.2 & 35.3k \\
  4$\times$2 & 50.0 & 55.9 & 43.0 & 49.6 & 5.9k & 52.0 & 50.5 &
  44.0 & 48.8 & 8.6k & 22.0 & 27.9 & 18.0 & 22.6 & 6.7k & 36.0 &
  41.1 & 32.8 & 36.6 & 9.3k \\
  4$\times$4 & 48.0 & 57.2 & 44.0 & 49.8 & 17.4k & 64.0 & 61.3 &
  49.0 & 58.1 & 24.0k & 17.0 & 24.6 & 17.0 & 19.5 & 20.4k & 37.6 &
  43.9 & 36.8 & 39.4 & 25.9k \\
  4$\times$6 & 50.0 & 58.7 & 43.0 & 50.6 & 31.7k & 59.0 & 56.0 &
  47.0 & 54.0 & 40.7k & 21.0 & 30.9 & 18.0 & 23.3 & 37.2k & 35.5 &
  44.2 & 31.4 & 37.0 & 45.3k \\
  4$\times$8 & 48.0 & 55.1 & 37.0 & 46.7 & 62.4k & 56.0 & 53.9 &
  44.0 & 51.3 & 53.4k & 20.0 & 26.6 & 18.0 & 21.5 & 72.9k & 34.4 &
  39.0 & 31.2 & 34.9 & 66.3k \\
  \bottomrule
  \end{tabularx}
\end{table*}
Table~\ref{tab:width-depth-results} presents the full grid search results across different width ($W$) and depth ($D$) configurations using the Qwen2.5-7B-Instruct model on multi-hop QA datasets. For the efficiency of the grid search, we report the result on a 100-sample subset of the dev set. For each configuration, we report accuracy, F1 score, exact match (EM), average score, and token usage across four multi-hop QA datasets. These results inform our selection of $W{=}2$ as the default width setting, which provides the best trade-off between performance and computational cost.

\section{Prompts for the LLM Agents}

\vspace{20pt}
\begin{center}
\begin{promptbox}{Context Manager Prompt}
Act as the context manager for a Retrieval-Augmented Generation (RAG) system. Your job is to maintain a single, up-to-date note that contains all the information relevant to answering the original query. Please ensure that the note includes all original text information useful for answering the question.

Steps:
- Based on the retrieved documents, supplement the notes with content not yet included but useful for answering the question.
- Resolve conflicts: if statements disagree, keep the most reliable or recent version.

End your response with the literal tag \texttt{[END]}.

Original query: \{query\}

Old note: \{note\}

New information: \{new\_context\}

Updated note:
\end{promptbox}
\end{center}

\vspace{6pt}
\begin{center}
\begin{promptbox}{Answer Generator Prompt}
Answer the question based on the given notes. 
Output ONLY the exact answer in as few words as possible. 
Do not include the question, reasoning, or any extra text. 
End your response with the literal tag [END].

The following are given notes: 
\{note\}

Question: \{query\}
Answer:
\end{promptbox}
\end{center}

\vspace{6pt}
\begin{center}
\begin{promptbox}{Multi-Path Dispatch Rewrite Prompt}
You are an intelligent assistant in a Retrieval-Augmented Generation (RAG) system. Your goal is to (a) diagnose retrieval needs for the current question and (b) produce exactly \{N\} rewritten queries, each paired with the most suitable retrieval strategy for that specific rewrite.

Information:
- Original Query: \{query\}
- Current Query: \{current\_query\}
- Context: \{context\}

Available Retrieval Strategies:
- bm25 (sparse / lexical): Prioritizes exact token and phrase matches.
- dense (semantic / vector similarity): Matches by meaning despite paraphrases.

Instructions:
1. Use the context to reflect on what is missing to answer the query. Think about both the big picture and the small atomic facts that might need verification.
2. Generate exactly \{N\} rewritten queries.
   - Do not just paraphrase — each query should explore a different angle, granularity, or fact.
   - Avoid near-duplicates.
   - Each query must serve a distinct retrieval purpose.
3. For each query, select the most suitable retrieval strategy:
   - Use bm25 when exact names, phrases, or quoted terms matter; remove ``wh'' words.
   - Use dense when searching for meanings, definitions, or related concepts.

Output Format (strict):
1. First provide your analysis and rationale in a <think> block, including per-item strategy justification.
2. Then output exactly \{N\} query rewrites using the following structure:
<queries>
  <item rank="1"><strategy>bm25|dense</strategy>
    <query>...</query>
  </item>
  <item rank="2"><strategy>bm25|dense</strategy>
    <query>...</query>
  </item>
  ...
  <item rank="\{N\}"><strategy>bm25|dense</strategy>
    <query>...</query>
  </item>
</queries>
3. End your response with the literal tag [END].

Output:
\end{promptbox}
\end{center}

\vspace{6pt}
\begin{center}
\begin{promptbox}{Answer Selection Prompt (\agent{Context Manager})}
Question: \{question\}

All Generated Answers:
\{answer\_blocks\}

Based on all the answers and reasoning provided, select the best answer. Consider accuracy and relevance to the question. Give the final answer directly. Then, provide a direct, concise, and accurate answer inside <answer> </answer> tags. End your response with the literal tag [END].

Final answer:
\end{promptbox}
\end{center}

\vspace{6pt}
\begin{center}
\begin{promptbox}{Context Merging Prompt (\agent{Context Manager})}
You are an expert at combining contexts in a Retrieval Augmented Generation system for answering question \{question\}. Here are several notes: \{reasoning\_list\}.
Combine the above notes into a single note that includes all information and is useful for final answer generation. Please ensure that the note includes all original text information useful for answering the question. End your response with the literal tag [END].
Your response:
\end{promptbox}
\end{center}

}

\end{document}